\documentclass[preprint,review,12pt]{elsarticle}

\usepackage{amssymb}

\usepackage{lineno}
\usepackage[colorlinks = true, linkcolor = blue, urlcolor = blue, citecolor = blue, anchorcolor = blue]{hyperref}
\modulolinenumbers[5]
\usepackage{todonotes}
\usepackage{pgfplots} 
\usepackage{graphicx}
\usepackage{changepage}

\usepackage{amsmath,amssymb,amsfonts}
\usepackage[capitalise,nameinlink]{cleveref}

\crefname{section}{Sect.}{Sect.}
\Crefname{section}{Section}{Sections}
\crefname{listing}{\lstlistingname}{\lstlistingname}
\Crefname{listing}{Listing}{Listings}

\usepackage{algorithmic}
\usepackage{floatrow}
\usepackage{balance}
\usepackage{calc}
\usepackage{textcomp}
\usepackage{xcolor}
\usepackage{subfig}
\usepackage{siunitx}

\usepackage{booktabs}
\usepackage[resetlabels,labeled]{multibib}
\newcites{S}{Selected Papers}
\usepackage{geometry}
\usepackage[normalem]{ulem}
\useunder{\uline}{\ul}{}
\usepackage{longtable}
\usepackage{pgf-pie}
\usepackage{bchart}
\usepackage{pifont}
\usepackage{caption}
\usepackage{subcaption}
\usepackage{comment}
\usepackage{setspace}
\usepackage{listings}

\usepackage{xcolor}
\usepackage{colortbl}

\usepackage{review-style}
\usepackage[enable]{modified}

\usepackage{glossaries}
\glsdisablehyper

\newacronym{SAS}{SAS}{self-adaptive system}
\newacronym{AFT}{AFT}{Attack-Fault Tree}
\newacronym{AT}{AT}{Attack Tree}
\newacronym{FT}{FT}{Fault Tree}
\newacronym{DFT}{DFT}{Dynamic Fault Tree}
\newacronym{PMC}{PMC}{probabilistic model checker}
\newacronym{SAFT-GT}{SAFT-GT}{SafeSec Attack-Fault Tree Generation Toolchain}
\newacronym{CVE}{CVE}{Common Vulnerabilities and Exposures}
\newacronym{CWE}{CWE}{Common Weakness Enumeration}

\newcommand{\textbox}[1]{
    \noindent\fbox{%
        \parbox{0.97\columnwidth}{%
            {#1}
        }%
    }
}
\definecolor{keywordcolor}{RGB}{116,20,83}
\definecolor{stringcolor}{RGB}{58,26,246}
\lstdefinestyle{SaftModelStyle}{
  language=Java,
  numbers=left,
  basicstyle=\ttfamily\footnotesize,
  stepnumber=1,
  numbersep=10pt,
  tabsize=2,
  showspaces=false,
  showstringspaces=false,
  numberstyle=\tiny\color{gray},
  keywordstyle=\bfseries\color{keywordcolor},
  morekeywords={Hazard,OR,BasicEvent,IntermediateEvent,AttackEvent, description,probability,dataflowElement,CVSSREQ,AND,AttackStep,deploymentElement,Component,Channel,property,Connect,version,frequency,RefComponent,Platform,executes,hostname,reachable,sshUser,toplevel,or,seq,lambda,and, ExploitabilityScore, ImpactScore, BaseScore, CVSS, CVE},
  stringstyle=\color{stringcolor},
}

\usepackage{ifthen}
\usepackage[normalem]{ulem} 
\newboolean{showcomments}
\setboolean{showcomments}{true} 
\ifthenelse{\boolean{showcomments}}
{
	\usepackage{mparhack}

	\newcommand{\nbbC}[3]{
		\marginpar[\hspace*{0.75cm}\parbox{35pt}{\tiny#1}]{\parbox{35pt}{\tiny#1}}
		{#2}
	}
	
	\newcommand{\removedC}[1]{{\color{red!90!black}\sout{#1}}}

	\newcommand{\rremovedC}[2]{\nbb{#1}{\color{red!90!black} \sout{#2}}{red!90!black}}

}
{
	\newcommand{\nbbC}[3]{}

	\newcommand{\removedC}[1]{}

	\newcommand{\rremovedC}[2]{}

}

\pgfplotsset{compat=1.18}

\newif\ifshowreview
\showreviewtrue

\journal{Journal of Systems and Software}

\begin{document}

\ifshowreview
\pagenumbering{roman}%
\pagenumbering{arabic}
\fi

\begin{frontmatter}


\title{Bridging Safety and Security in Complex Systems: A Model-Based Approach with SAFT-GT Toolchain}


\author[labela]{Irdin Pekaric}\corref{author1}
\cortext[author1]{Corresponding author}
\author[label6]{Raffaela Groner}
\author[label2]{Alexander Raschke}
\author[label2]{Thomas Witte}
\author[label1]{Jubril Gbolahan Adigun}
\author[label1,label4,label5]{Michael Felderer}
\author[label2]{Matthias Tichy}

\address[labela]{irdin.pekaric@uni.li \\
    Universität Liechtenstein \\
    Department of Information Systems and Computer Science \\
     Fürst-Franz-Josef-Strasse, 9490 Vaduz, Liechtenstein
}

\address[label6]{raffaela@chalmers.se \\
    Chalmers University of Technology and University of Gothenburg \\
    Department of Computer Science and Engineering\\
    H\"{o}rselg\r{a}ngen, 41296 Gothenburg, Sweden
}

\address[label2]{Firstname.Lastname@uni-ulm.de \\
    Ulm University \\
    Institute of Software Engineering and Programming Languages \\
    James-Franck-Ring 9, 89081 Ulm, Germany
}

\address[label1]{Firstname.Lastname@uibk.ac.at \\
    University of Innsbruck \\
    Department of Computer Science \\
    Technikerstraße 21a, A-6020 Innsbruck, Austria
}

\address[label4]{michael.felderer@dlr.de \\
    German Aerospace Center (DLR) \\
    Institute of Software Technology \\
    Muenchener Strasse 20, 82234 Wessling, Germany
}

\address[label5]{University of Cologne \\
    Department of Mathematics and Computer Science \\
    Albertus-Magnus-Platz, 50923 Cologne, Germany
}

\begin{abstract}
In the rapidly evolving landscape of software engineering, the demand for robust and secure systems has become increasingly critical. This is especially true for self-adaptive systems due to their complexity and the dynamic environments in which they operate. To address this issue, we designed and developed the SAFT-GT toolchain that tackles the multifaceted challenges associated with ensuring both safety and security. This paper provides a comprehensive description of the toolchain's architecture and functionalities, including the Attack-Fault Trees generation and model combination approaches. We emphasize the toolchain's ability to integrate seamlessly with existing systems, allowing for enhanced safety and security analyses without requiring extensive modifications and domain knowledge. Our proposed approach can address evolving security threats, including both known vulnerabilities and emerging attack vectors that could compromise the system. As a use case for the toolchain, we integrate it into the feedback loop of self-adaptive systems. Finally, to validate the practical applicability of the toolchain, we conducted an extensive user study involving domain experts, whose insights and feedback underscore the toolchain's relevance and usability in real-world scenarios. Our findings demonstrate the toolchain's effectiveness in real-world applications while highlighting areas for future improvements. The toolchain and associated resources are available in an open-source repository to promote reproducibility and encourage further research in this field.
\end{abstract}

\begin{highlights}
\item The SAFT-GT toolchain enables semi-automatic Attack-Fault Tree generation for enhanced safety and security assessment in self-adaptive systems.

\item The toolchain efficiently integrates into the feedback loop of self-adaptive systems, allowing for dynamic updates based on security assessments.

\item A user study with domain experts confirms the toolchain's relevance and practical applicability in real-world scenarios.

\item Performance experiments demonstrate that the Attack-Fault Tree generation pipeline operates within feasible time constraints, supporting real-time applications.

\item The complete toolchain and resources are provided for download, fostering further research and collaboration in the field. 
\end{highlights}

\begin{keyword}
attack-fault tree \sep safety and security analysis \sep self-adaptive system \sep model formalism \sep expert survey
\end{keyword}

\end{frontmatter}


\section{Introduction}  
\label{sec:introduction}

The growing complexities of modern systems have necessitated the employment of different approaches in their analysis~\cite{SLR}. These approaches often attempt to reduce human effort in the delivery of various services. This is achieved through the incorporation of autonomy and allowing systems to be aware of their physical environment \cite{ceccarelli2023evaluating}. This fundamental ability enables software systems to sense and reason about the system behaviors as well as to react by exhibiting self-adaptation --- giving rise to \glspl{SAS}. \glspl{SAS} can modify and optimize their performance based on feedback and the changing conditions, as well as the system goals \cite{10.1145/3589227}. 

Despite all the benefits that \glspl{SAS} provide, they also bring various challenges that often involve security and safety concerns. According to the practitioners in the \glspl{SAS} domain, these two aspects represent some of the biggest problems related to these types of systems. This was demonstrated in the recent industrial survey conducted by Weyns et al. \cite{10.1145/3589227}. The issues often occur due to various types of vulnerabilities, which can cause parts of a system to fail, causing the overall state of a system to become unsafe. Many of these vulnerabilities arise from diverse hardware and software configurations, which necessitates an ongoing analysis and vigilant monitoring throughout their operational lifecycle. Given the interconnected nature of these challenges, it is of utmost importance to consider security and safety jointly, as these two aspects significantly affect each other and can determine the overall reliability of the system.
 
However, joint consideration of safety and security aspects in the context of \glspl{SAS} and their uncertainty is a complex and challenging endeavor~\cite{10.1145/3503229.3547048}.
One promising approach to handle this complexity is to leverage a joint safety and security analysis using \glspl{AFT}. \glspl{AFT} are a powerful analytical tool that combines \glspl{AT} and \glspl{FT} in a single modeling formalism~\cite{Stoelinga2,Stoelinga1}, allowing for a systematic exploration of the interplay between safety and security issues~\cite{groner2023model}. They provide a graphical representation of the various ways in which a system can fail (faults) or be compromised (through attacks~\cite{vulner,attacks}), and how these faults and attacks can propagate through the system. This is achieved by integrating safety and security considerations at runtime when the system is in its operational context, allowing for a more holistic and robust analysis of potential safety risks and system vulnerabilities~\cite{10.1145/3524844.3528062}.  

\glspl{AFT} can be verified using existing model-checking techniques, e.g., critical path analysis, or calculation of failure rates and probabilities. Generating large \glspl{AFT} for realistic systems by hand, however, is error-prone and infeasible. In comparison, it is much easier for safety experts to create only \gls{FT} models for a target system separately as well as security experts generating \gls{AT} models, which are at a much lower abstraction level~\cite{ComponentBasedHazard} compared to creating \glspl{AFT}.

One disadvantage of \glspl{AFT}, however, is that they cannot model the system state but only possible hazards.
Thus, we restrict ourselves to architectural reconfigurations that change the data flow and/or the hardware and software components of a managed system.
Since our \gls{SAFT-GT} \cite{groner2023model} generates \glspl{AFT} for cyber-physical systems based on the data flow and the used hardware and software components, the \glspl{AFT} we analyze take the current system state into account implicitly. 

By incorporating \glspl{AFT} into the \glspl{SAS} loop, the system can leverage real-time data to dynamically update its models (due to potential security and safety issues), which corresponds to their defining characteristics. Furthermore, the generation of \glspl{AFT} allows for rapid adaptation to new threats and vulnerabilities, enhancing the system's resilience. This integration not only facilitates proactive risk management but also supports informed decision-making in the planning phase, ultimately leading to more robust and secure self-adaptive systems that can autonomously respond to emerging challenges in their operational environment.

\subsection{Overview of the Approach}{\label{sec:overview}}

\begin{figure*}[htb]
    \centering
    \includegraphics[width=\textwidth]{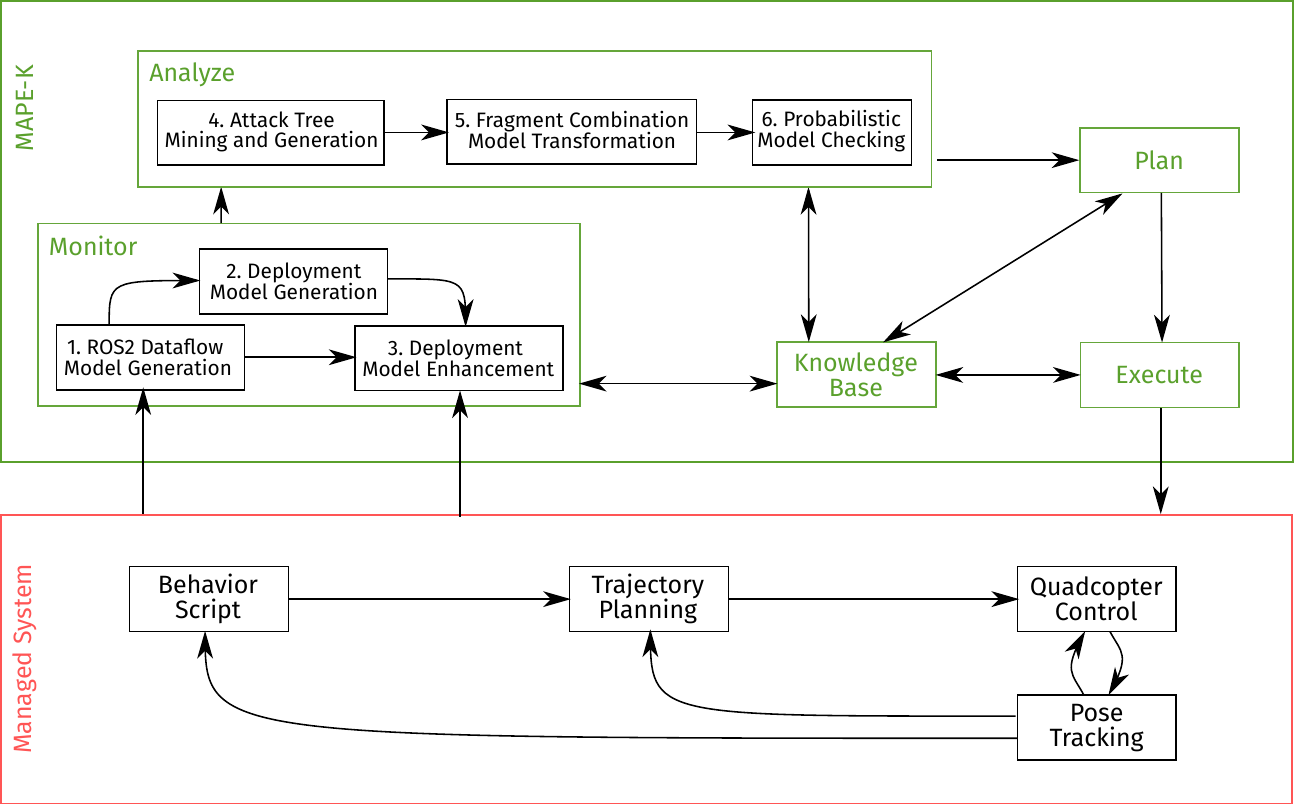}
    \caption{Integration of our approach in the MAPE-K loop.}
    \label{fig:mape-k}
\end{figure*}

Figure~\ref{fig:mape-k} shows the integration of the presented approach in a self-adaptive system controlled by a MAPE-K~\cite{1160055} loop: First, a managed system forms the basis. Our example system implements a control system for a quadcopter with ROS2 components (robot operating system~\cite{macenski2022robot}) to plan trajectories and avoid obstacles, as well as high-level scripting of robot behavior, e.\,g., waypoints to follow. Second, a MAPE-K subsystem that monitors the state of the ROS node graph and messages, can reconfigure the application by activating or deactivating components of the managed system, e.\,g., to switch between a mission and safety behavior, such as flying back to origin. Finally, our approach performs the safety and security analysis in the MAPE-K loop feeding the knowledge base with risk values for hazards. This information is used by the planning component to influence or trigger future adaptation decisions.
More details are given in Figure~\ref{fig:overview} in Section~\ref{sec:models}. 

Our approach focuses on the security problems that lead to a failure of the system. 
Therefore, other security aspects, such as confidentiality or data integrity, are only considered if they can also lead to a failure.
For example, the manipulation of sensor data might lead to a drone crashing into an obstacle due to manipulated LiDAR sensor data.


\subsection{Contributions} 
In order to close the feedback loop, we extend our previously published SAFT-GT~\cite{groner2023model} pipeline by a transformation of the generated \glspl{AFT} into \glspl{DFT} and use them as input for the Storm model checker\footnote{\url{https://www.stormchecker.org/}}.
We evaluate our extension and the overall approach by conducting a workshop with domain experts and an experimental evaluation of the approach.
Additionally, we perform time measurements to evaluate the performance of the extended \gls{SAFT-GT} pipeline, which provides insights on whether the pipeline is applicable in a real system.
Specifically, we provide the following contributions:

\begin{itemize}
    \item We provide a detailed description of our previously implemented \gls{SAFT-GT} approach \cite{groner2023model}.
    \item We extend our \gls{SAFT-GT} approach by applying a \gls{PMC} to verify the output of the \gls{AFT} generator.
    \item We link and evaluate \gls{SAFT-GT} through the feedback loop of self-adaptive systems.
    To this end, we used a \gls{SAS} built upon the robot operating system (ROS)~\cite{macenski2022robot}.
    \item We conduct a workshop that includes two surveys and expert discussions. Based on the insights we obtained, we discuss the strengths and limitations of the developed approach and outline future research directions. 
    \item We perform experiments to evaluate the performance of the proposed approach.
\end{itemize}

We provide the complete toolchain, including the models of the presented approach and examples for download, fostering further research and collaboration in the field: \url{https://github.com/sp-uulm/saft-gt}.

\subsection{Paper Structure}
The remainder of the paper is structured as follows: Section~\ref{sec:background} provides background information on self-adaptive systems, safety and security modeling and security concepts (including metrics). Section~\ref{sec:running_example} demonstrates a running example for which the proposed approach can be applied. Section~\ref{sec:models} presents each separate model that we utilize in our approach, while Section~\ref{sec:comibationOfModels} explains how these models are combined, verified and integrated into the feedback loop of self-adaptive systems. Section~\ref{sec:evaluation} offers an evaluation of the developed approach through expert surveys, discussions as well as performance evaluation. Section~\ref{sec:discussion} elaborates on the limitations of the approach as well as experts' feedback. Section~\ref{sec:related} discusses the related work on \gls{AFT} generation. Finally, Section~\ref{sec:conclusion} concludes the work.

\section{Background} \label{sec:background}
In this section, we provide the necessary background on self-adaptive systems, safety and security models used in this work and relevant security concepts and metrics.

\subsection{Self-adaptive System}\label{sec:sas}
\glspl{SAS} are systems that adjust their behavior to optimally fulfill their requirements even if their environment or goals change.
Usually, a \gls{SAS} consists of two subsystems: 1) the managed subsystem, which should fulfill a predefined goal, and 2) the managing subsystem that controls the adjustment of the behavior of the managed subsystem~\cite{Weyns2013}.
For example, a drone that should independently plan and execute missions can be considered as a part of the \gls{SAS}. The drone forms the managed subsystem, and the corresponding managing subsystem plans new missions and adjusts the drone's behavior to complete a mission.

A well-known architecture used for the managing subsystem is MAPE-K (\textbf{M}onitor, \textbf{A}nalyze, \textbf{P}lan, \textbf{E}xecute - \textbf{K}nowledge)~\cite{1160055}.
Figure~\ref{fig:mape-k} in Section~\ref{sec:overview} provides an overview of the relationships between managed subsystem, managing subsystem, and MAPE-K.
The Monitor component observes the state and behavior of the managed subsystem as well as its environment.
The Analyze component decides whether the managed subsystem should adapt its behavior.
For example, the analyze component might detect that the drone is running low on energy and decide to trigger an adaptation to avoid damage due to a crash.
The Plan component plans the adjustment of the managed subsystem.
For example, the Plan component determines a plan that forces the drone to abandon its current mission and initiate a safe landing.
Finally, the Execute component implements the plan of the Plan component.
The Knowledge component serves as the centralized provision of information.
The other components use the Knowledge component to store information and to communicate indirectly with each other~\cite{7194653}.

\subsection{Safety and Security Modeling}\label{sec:backgroundModelling}
There is a variety of modeling approaches used in the context of safety and security analysis. Our work leverages the security-related models \glspl{AT}, the safety-related models \glspl{FT}, and \glspl{AFT}, which are used in a joined safety and security modeling approach. 
In the following, we provide background information on these three established models in more detail. We use a running example that describes how a flooding attack can lead to a drone crashing into a person to illustrate these models.

\subsubsection{Attack Trees}\label{sec:backgroundattackTrees}
\glspl{AT} are models used to represent adversary actions regarding how a certain system or component can be targeted~\cite{mauw2006foundations,schneier1999modeling}. 
They consist of a root node, which represents the goal of an attack, attack steps or intermediate targets, and logical gates.
Depending on the goal of the analysis, different values are assigned to attack sets within the three. These can include values such as the expected costs or the probability of performing the respective attack set successfully.
Logical gates, e.g., AND and OR, are used to model the relationships between individual attack steps necessary to reach the goal of the modeled attack~\cite{mauw2006foundations,muzammil2024unveiling,lallie2020review}.

\begin{figure}[ht]
\centerline{\includegraphics[width=0.55\columnwidth]{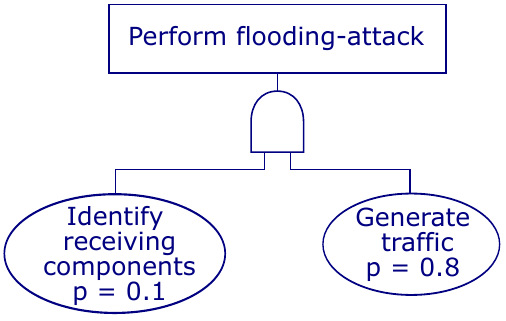}}
\caption{Example of an Attack Tree~(AT).}\label{fig:exampleAttackTree}
\end{figure}

Figure~\ref{fig:exampleAttackTree} shows a simplified example of an \gls{AT} for a packet flooding attack. It demonstrates that an attacker needs to  ``identify the receiving components'' in a system and ``generate traffic'' in order to perform a flooding attack.
The probabilities specified for each attack step~(depicted as ellipses) indicate the probability of successfully performing the respective attack step.

\subsubsection{Fault Trees}\label{sec:backgroundfaultTrees}
\glspl{FT} represent a common formalism from safety analysis that is used to model possible hazards and their causes~\cite{vesely1981fault,pai2002automatic}.
In addition to standard \glspl{FT}, there are also more specialized variants, such as \glspl{DFT}~\cite{DFTs}, which consider temporal sequences of events or Repairable Fault Trees can also model complex repairable systems~\cite{1311936}.
However, \glspl{FT} are in general acyclic graphs that consist of two categories of nodes, namely events and logical gates.
Regarding events, a distinction is drawn between the top event, basic events, and intermediate events.
The top event represents the hazard to be analyzed.
Basic events model events that occur randomly and intermediate events model events caused by other events.
Logical gates are used to model the propagation of failures through the system.
Standard \glspl{FT} can only model simple logical relationships such as AND and OR~\cite{Stoelinga3}. 
\glspl{DFT}, however, offer dynamic and temporal gates that enable more complex combinations of events.
For example, a Simultaneous-AND~(SAND) gate triggers when all incoming events occur at the same time.
Another example is a Priority-AND~(PAND) gate, which triggers when all incoming events occur in their predefined order~\cite{10.1007/978-3-642-33678-2_9}.

\begin{figure}[ht]
\centerline{\includegraphics[width=0.6\columnwidth]{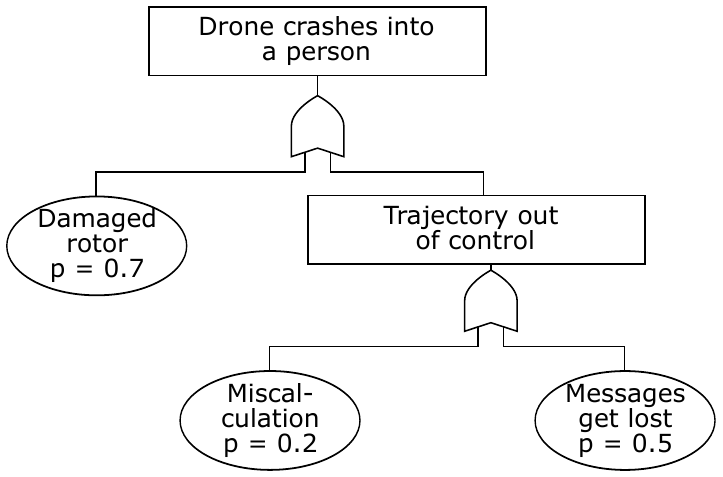}}
\caption{Example of a Fault Tree~(FT).}\label{fig:exampleFaultTree}
\end{figure}

Figure~\ref{fig:exampleFaultTree} shows a simplified example of a \gls{FT}.
This \gls{FT} models that the failure that ``Drone crashes into a person'' can occur due to a damaged rotor with a probability of 0.7 or due to the drone being out of control, which can be caused by a miscalculation with a probability of 0.2 or by losing messages with a probability of 0.5.

\subsubsection{Attack-Fault Trees}\label{sec:backgroundattackfaultTrees}

Kumar and Stoelinga~\cite{ Stoelinga1} present in their work Attack-Fault Trees~(\gls{AFT}), which combines \glspl{FT} with \glspl{AT} into a unified modeling mechanism to leverage a joint analysis of safety and security.
To this end, \glspl{FT} are enhanced with additional events that represent faults caused by malicious actions.
These events are also called attack events.
Attack events are analogous to an attack goal that serves as the root of an \gls{AT} and, therefore, mark the juncture between \gls{FT} and \gls{AT} in an \gls{AFT}~\cite{Stoelinga3,AFTsOrig}.

\begin{figure}[ht]
\centerline{\includegraphics[width=0.6\columnwidth]{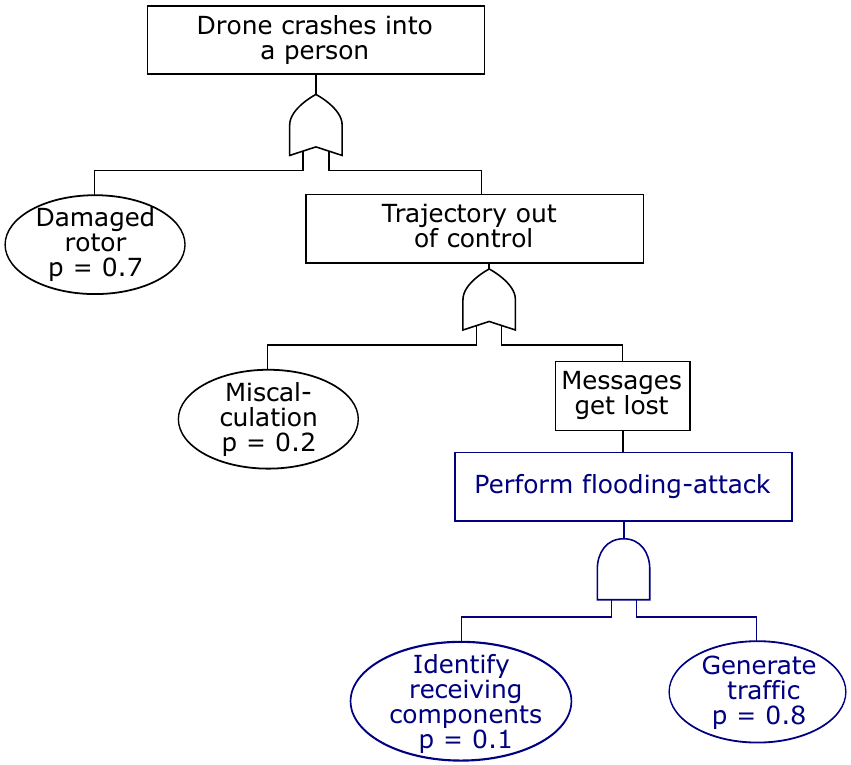}}
\caption{Example of an Attack-Fault Tree~(AFT).}\label{fig:exampleAttackFaultTree}
\end{figure}

Figure~\ref{fig:exampleAttackFaultTree} shows an \gls{AFT} resulting from combining the \gls{FT} in Figure~\ref{fig:exampleFaultTree} and the \gls{AT} in Figure~\ref{fig:exampleAttackTree}.
It should be noted that our example \gls{FT} has been extended by one intermediate event (``Messages get lost'') to model the context in which an attack leads to failure.

This example \gls{AFT} describes possible events that can cause a drone to crash into a person.
On the one hand, a damaged rotor can cause this failure.
On the other hand, the loss of control over the trajectory of the drone can lead to the failure mentioned above.
A miscalculation or the loss of messages can lead to the trajectory being out of control.
Since the loss of messages can be caused by performing a flooding attack, the \gls{FT} part of the \gls{AFT} is extended with the \gls{AT} in Figure~\ref{fig:exampleAttackTree}.

\subsection{Security Concepts and Metrics}\label{sec:background_metrics}

\noindent In this section, we outline the general security concepts and metrics that were utilized in the proposed modeling approach. CVE\footnote{Common Vulnerabilities and Exposures, \url{https://cve.mitre.org/}} data is used to distinguish between different \textit{vulnerabilities}. These vulnerabilities have already been identified by security experts and recorded in order to provide a means of protection against existing threats. Each CVE has a unique identifier, description, the links to technical reports, advisories and patches.

CWE\footnote{Common Weakness Enumeration, \url{https://cwe.mitre.org/}} entries represent specific higher-level groups to which CVEs are associated. This is done in order to provide a hierarchy for vulnerability data and these are formulated as \textit{weaknesses}. There are two main hierarchies, which divide CVEs into software and hardware weakness types. Most of the CVEs contain a CVSS\footnote{Common Vulnerability Scoring System, \url{https://www.first.org/cvss/}} vector. It provides various types of qualitative scores related to the severity of the vulnerability. Besides providing more detailed information for assessing severity, it also includes information about the impact of a CVE, the so-called CIA triad \cite{samonas2014cia, sauerwein2019analysis}.

CPEs\footnote{Common Platform Enumerations, \url{https://nvd.nist.gov/products/cpe}} portray various systems, software, and platforms, which are represented using syntax for Uniform Resource Identifiers (URI) including a specific version of a software or library. This allows security engineers and researchers to know exactly which software is affected by a certain CVE. In cybersecurity, attacks can also occur by exploiting multiple vulnerabilities, which form attack chains. Some of these chains are similar because they address CVEs that belong to the same CWE or they have corresponding mechanics. To represent these similarities, CAPEC\footnote{Common Attack Pattern Enumeration and Classification, \url{https://capec.mitre.org/}} entries were created, which allow an easy understanding of common attacker actions. The aforementioned databases present a solid foundation for the proposed approach since they include vulnerabilities, weaknesses, platforms, and attack patterns.

Lastly, we also utilize the ``Exploit Prediction Scoring System'' (EPSS\footnote{Exploit Prediction Scoring System, \url{https://www.first.org/epss}}). It describes ``the probability that a vulnerability will be exploited in the wild within the first 12 months after public disclosure''\cite{epss2021}. Since the existence of a system vulnerability does not mean that the vulnerability is instantly exploited, the score is used to estimate the probability of a successful attack over time (generated as a part of the attack model, wherein the scores are assigned automatically). The EPSS score is applied because we do not consider specific attacker profiles or other more detailed aspects, such as the duration and the success of an attack. Thus, an exponential distribution can be assumed for an \gls{AFT}~\cite{Stoelinga1}, which is also the default approach for \gls{FT} analysis~\cite{Stoelinga3}. The reason we do not consider attacker profiles is that we aim to make our approach as automated as possible, whereas attacker profiles need to be manually crafted by security experts. In addition, this can also result in a false sense of security due to profile inaccuracies as well as raise potential scalability issues \cite{lenin2014attacker}. However, due to the flexibility of the toolset, manual specification and integration of attacker profiles in the generated attack models are still possible.

\section{Running Example}  
\label{sec:running_example}
We introduce a small example that we will use in the following sections to illustrate the models used. The scenario is an automatically controlled drone that could be used, for instance, for reconnaissance flights in agriculture to detect hidden animals in a field before the crops are harvested. The drone receives the trajectory from a control computer and transmits its position back so that the control computer can adjust the trajectory multiple times. We have simulated this scenario in our drone laboratory with the following necessary adjustments: The position data of the drone is determined via an external camera system (Optitrack). This data is then transferred to the actual control computer, which calculates the trajectory of the drone according to the area to be covered. The drone is then gradually sent instructions via the Wi-Fi network as to which points to fly over one after the other. Our system is based on ROS and consists of several nodes that communicate and interact with each other and perform various tasks such as transmitting position data, calculating the trajectory or communicating with the drone. 

In the following, we will only consider the risk of a drone flying into a person. The manually developed \gls{FT} therefore includes intermediate events such as ``Trajectory out of control'' or ``Drone unable to fly''. These events can occur if the drone's hardware has problems (broken rotors, low battery, etc.), but also if the control messages get lost. The latter can be due to poor connection quality as well as to specific attacks targeting the communication channel or the control computer directly. The individual nodes could, for example, be flooded with messages as a result of a denial-of-service (DoS) attack and thus no longer be able to keep up with the calculation. In the example \gls{FT}, attack events targeting precisely those nodes that supply data for the node calculating the trajectory are therefore used as basic events.

While this is a simple example, it effectively demonstrates the potential of our approach. All models presented in the remainder of the paper, as well as the evaluation, are based on this example. 

\section{Models}  
\label{sec:models}
In this section, we describe the use, the textual representation, and the process of obtaining each model we utilize in our AFT generation toolchain. 
We provide a detailed description of our generation approach that relies on these models in Section~\ref{sec:comibationOfModels}.

\begin{figure*}[htb]
    \centering
    \includegraphics[width=\textwidth]{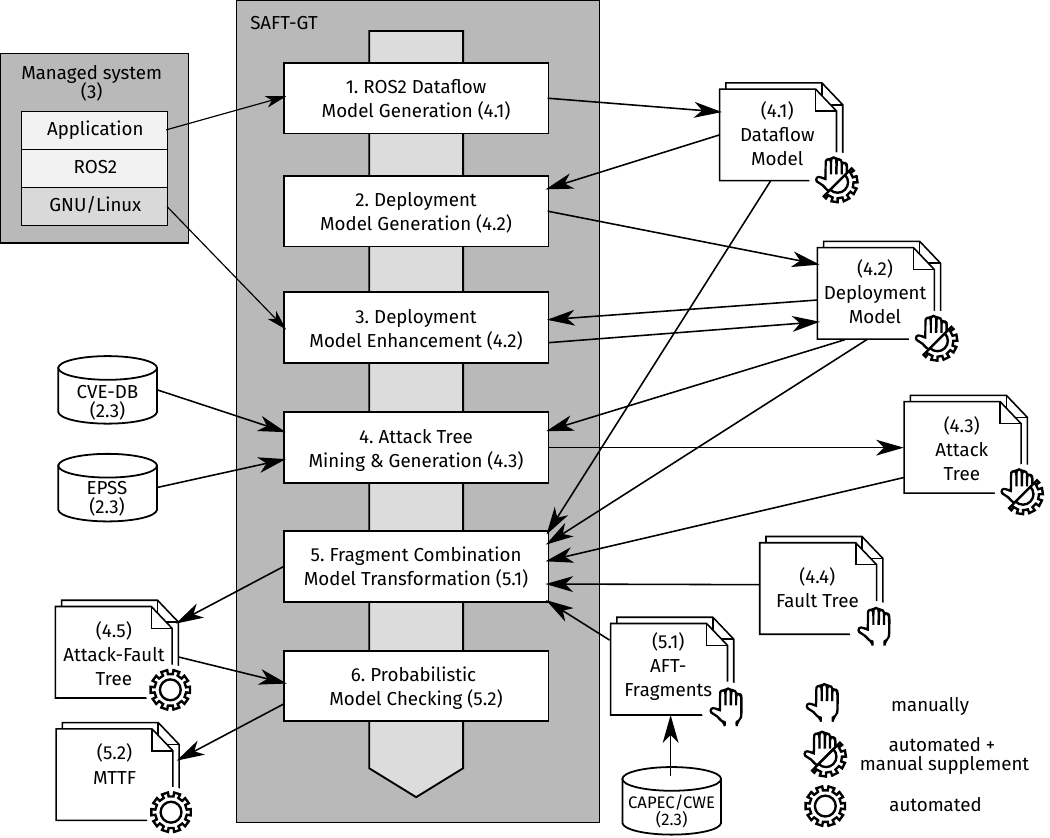}
    \caption{Overview of the SAFT toolchain and produced/used artifacts.}
    \label{fig:overview}
\end{figure*}

Figure~\ref{fig:overview} includes all the individual steps and involved models of the presented toolchain. Detailed information about how these models are obtained is given in the subsections indicated in the figure. The principal run is as follows: After a reconfiguration of the ROS2 system, a run of the toolchain is triggered in the MAPE-K loop. This run starts with the (re-)generation of the (adapted) dataflow between the ROS nodes. From this dataflow model, an initial deployment model is derived that is enhanced with dependency information extracted from the running system. In the next step, open-source security databases (see Section~\ref{sec:background_metrics}) are mined in order to identify specific vulnerabilities that affect components used in the system. For each vulnerable component, an \gls{AT} is generated. These \glspl{AT} are combined with manually created and provided \glspl{FT} using so-called \gls{AFT}-fragments that are manually created by security and modeling experts. The resulting \gls{AFT} is translated into an input model for a \gls{PMC} whose result (MTTF) is fed back to the Knowledge base.

\subsection{Dataflow Model}
\noindent\textbf{Summary:} 
Dataflow represents the dependency between two components. For example, if component B receives messages from component A, then there is a dependency between component B and component A. As a consequence, if component A has been compromised, it is easy to manipulate component B, so component B should also be considered compromised. In order to capture the consequences of an attack, it is therefore important in our approach to model the data flow between compromised components. 

The dataflow model is a simplification of the ROS component meta-model similar to the abstract component meta-model in \cite{gherardi2013variability}. The model is used to generate and integrate dataflow models from any system. 
It captures the logical components of a system and dataflows between them. This logical view is more abstract from the actual implementation and focuses on the communication between components. In consequence, the meta-model only consists of \emph{components} -- entities that provide, transform or process data (such as sensors, actors, or data processing units) -- and \emph{channels} -- communication paths between components to send or receive messages, call functions, etc.

Since our prototypical implementation builds upon a ROS system, we provide a dataflow model generator for ROS2 systems where ROS nodes are mapped to components while ROS topics, services, and actions are mapped to channels. To overcome the problem that ROS2 systems do not define their interface statically, but connect to topics and services dynamically, our generator consists of a single ROS node that can be triggered to collect architectural and dataflow data using ROS' introspection capabilities at runtime. The generator can be triggered repeatedly to monitor the system for changes or architectural reconfigurations. 

Additionally, the textual representation of the model is designed to be manually extendable: additional components and channels can be defined manually and incorporated with the rest of the model.\\
\noindent\textbf{Obtained:} Automatically extracted from a running system and/or manually extended.\\

\lstset{style=SaftModelStyle}
\begin{lstlisting}[caption={Excerpt of the dataflow model of our running example.},label=lst:dataflowListing]
Component simple_trajectory_server {
    property ros_name = "simple_trajectory_server";
}

Component trajectory_client {
    property full_ros_name = "/trajectory_client";
}

Channel trajectory_assign {
    property full_ros_name = "/trajectory_assignment";
    property ros_type = "msg/SeTrajectoryAssignment";
    property ros_channel_type = "topic";
}
Connect Component=simple_trajectory_server -> Channel=trajectory_assign; 
Connect Component=trajectory_client <- Channel=trajectory_assign;
\end{lstlisting}

\noindent\textbf{Textual Representation:}
Listing~\ref{lst:dataflowListing} shows a small excerpt of the dataflow model of our running example, which defines two components \textit{simple\_trajectory\_server} and \textit{trajectory\_client}, and a channel \textit{trajectory\_assign}. Arbitrary properties can be added to components and channels by providing more detailed information about the corresponding entity. A component can be connected to a channel via the keyword \texttt{Connect}, indicating the direction of communication wherein \texttt{->} signifies outgoing, while \texttt{<-} represents incoming messages. This separate description of the communication ends makes it possible to define hyperedges with multiple senders and/or multiple recipients of a message, which is also possible in ROS2.\break

\subsection{Deployment Model}
\label{sec:deployment_model}
\noindent\textbf{Summary:} 
The deployment model denotes how the components and channels from the dataflow model are deployed on a specific system. This information has to be provided manually. Our toolchain automatically extends this initial information with the libraries on which a component depends. Dependencies that cannot be derived automatically (e.g., because the platform that a component runs on cannot be reached by our analysis tool) can be added manually. We do not rely on the component's source code to obtain dependency information as snyk\footnote{\url{https://snyk.io}}. On the contrary, we utilize information about open files of the running processes of a target component. Since we directly identify all libraries and components present on the running system, including their specific versions, we intentionally do not model countermeasures explicitly. Instead, we focus on currently vulnerable components as reported by vulnerability databases. If a component has been patched or a countermeasure has been applied effectively, the corresponding version will no longer be marked as vulnerable and is therefore excluded from our analysis. This approach ensures that only relevant vulnerabilities are considered.

\begin{figure}
    \centering
    \includegraphics[width=\textwidth-3cm]{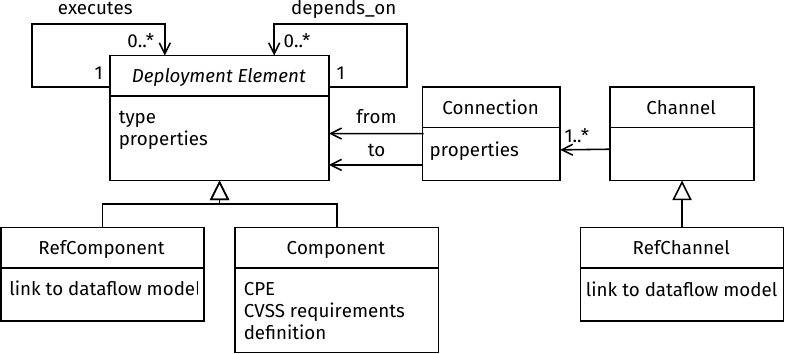}
    \caption{Meta-model of deployment model.}
    \label{fig:deploymentmodel}
\end{figure}

Fig.~\ref{fig:deploymentmodel} shows a simplified meta-model of the deployment model. A \textit{Deployment Element} is either a newly defined \textit{Component} or a reference to a dataflow component (\textit{RefComponent}). Similarly, new \textit{Channel}s between \textit{Deployment Element}s can be defined --- in addition to the already defined \textit{Channel}s within the dataflow model (\textit{RefChannel}). Each component has an optional type representing different abstraction levels (e.g., File, Library, Package, Platform, OS, etc.) and arbitrary properties (key/value pairs). A (low-level) component might be related to a CPE entry or have CVSS requirements.  

Each deployment element can be \textit{executed} on another element or \textit{depend on} other elements. This dependency information is generated recursively by our analysis tool via the used files and libraries of a component returned by Unix tools such as \texttt{lsof}\footnote{\url{https://github.com/lsof-org/lsof}} and \texttt{ldd}\footnote{\url{https://linux.die.net/man/1/ldd}}. System-specific package managers as \texttt{apt}\footnote{\url{https://wiki.debian.org/AptCLI}} and \texttt{dpkg}\footnote{\url{https://wiki.debian.org/Teams/Dpkg/}} abstract this information into package names for which CVEs can be identified. So far, the tool supports Ubuntu and Gentoo as platforms, but its architecture includes several abstraction layers to facilitate the integration of other platforms.

In the following step, the goal is to find the corresponding CPE for each identified package. For this purpose, we use the tool CPEguesser\footnote{\url{https://github.com/cve-search/cpe-guesser}} in combination with some heuristic preprocessing such as shortening names, removing additional version information, etc. Lists, CPEs and all packages for which no CPE entry can be found are then passed to the \gls{AT} Generator to determine possible CVEs for the identified software libraries. 

\noindent\textbf{Obtained:} Automatically extracted from a running system and/or manually.

\noindent\textbf{Textual Representation:} 
In our running example, most ROS nodes run on a \textit{Platform} called ``rosbox''. The properties of the ``rosbox'', including information about how our analysis tool can reach it, are used to gather more detailed information about the components running on it (see the excerpt in Listing~\ref{lst:deploymentListing}). In the listing, one can see the already discovered dependency of the component ``simple\_trajectory\_server'' for library ``libfastrtps'' version 2.1.1.
\lstset{style=SaftModelStyle}
\begin{lstlisting}[caption={Excerpt of the deployment model of our running example.},label=lst:deploymentListing]
ros_nodes:Platform = {simple_trajectory_server, quad_state_node, 
                      bebop_driver_node, optitrack_motive, ...}

rosbox (hostname="x.x.x.x",reachable="ssh",sshUser="ros") = 
        {ROS_foxy, Ubuntu_20, Linux}
rosbox executes {ros_nodes}

RefComponent simple_trajectory_server 
    (ros_name="simple_trajectory_server") =
        {rosidl_typesupport_fastrtps_cpp_so, libfastcdr_so_1_0_13, 
         libfastrtps_so_2_1_1, ...}
\end{lstlisting}

\subsection{Attack Tree Model}\label{sec:attackTreeModel}

\noindent\textbf{Summary:} The Attack Tree model is utilized to represent potential adversarial actions that can compromise the security of a system. It systematically outlines how an attacker can achieve specific goals by exploiting vulnerabilities within the system. Each node in the tree represents an attack step or a target, while the edges illustrate the relationships and dependencies between these steps. This hierarchical structure allows for a clear visualization of the various paths an attacker might take to reach their objective.

To effectively model adversarial actions, we define attack events within the \gls{AT} that represent specific attack steps or exploits. Each attack event is characterized by its targeted software library (CPE), the method of attack (CVE), and the conditions that must be met for the attack to be successful. The \gls{AT} employs logical gates to illustrate the relationships and dependencies between various attack steps, enabling a nuanced understanding of how different attack paths can converge or diverge.

For instance, in an \gls{AT}, multiple attack events may be represented to illustrate different ways to compromise a system. One attack path could specify that to disable the communication channel of a drone, an attacker could flood the system with an extremely high number of network packets (DoS). Another attack scenario might involve exploiting a software vulnerability to gain unauthorized access to the drone's control system. These relationships can be represented using logical gates to show that both conditions must be satisfied for the attack to succeed. By analyzing these attack paths, security experts can prioritize defenses and mitigation strategies based on the likelihood and impact of various attack scenarios. More information on how attack models are generated can be found as part of our previous work \cite{pekaric2023streamlining}.

\noindent\textbf{Obtained:} Automatically generated from a running system based on library and package information.

\noindent\textbf{Textual Representation:} Listing \ref{attackTreeListing} demonstrates an example of the notation that we utilize for \glspl{AT}. Each attack is presented as an~\texttt{\color{keywordcolor}{IntermediateEvent}}, which is generated for each identified CPE of the running system. The CPE itself is stated in the description of the~\texttt{\color{keywordcolor}{IntermediateEvent}}. Since each CPE can be affected by multiple CVEs, these are presented as separate~\texttt{\color{keywordcolor}{AttackSteps}} or nodes that an attacker can exploit. ~\texttt{\color{keywordcolor}{AttackSteps}} are connected to the~\texttt{\color{keywordcolor}{IntermediateEvent}} by logical gates\footnote{For clarity and consistency, only standard logical gates (AND, OR, PAND, SAND) are supported in the current implementation, since the semantics of dynamic or repair gates (e.g., FDEP, SPARE) remain controversial when combining safety and security aspects.} (AND, OR, PAND, etc.). More sophisticated attacks can involve additional child nodes that are linked using more complex logical gates, such as AND or PAND gates wherein an attacker must exploit vulnerability A before vulnerability B. Each ~\texttt{\color{keywordcolor}{AttackStep}} also contains CVE's~\texttt{\color{keywordcolor}{description}},~\texttt{\color{keywordcolor}{CVSS}} vector,~\texttt{\color{keywordcolor}{BaseScore}},~\texttt{\color{keywordcolor}{ImpactScore}} and~\texttt{\color{keywordcolor}{Ex\-ploi\-ta\-bi\-li\-ty\-Score}}.

\lstset{style=SaftModelStyle}
\begin{lstlisting}[caption={Textual representation of an Attack Tree}, label=attackTreeListing]
IntermediateEvent description = "Generated for search by cpe for
    keyword: cpe:2.3:a:eprosima:fast_dds:2.1.1 Insufficient Control
    of Network Message Volume (Network Amplification)" {
    OR {
         AttackStep  CVE202138425  
         description = "eProsima Fast DDS
            versions prior to 2.4.0 (#2269) are susceptible to
            exploitation when an attacker sends a specially crafted
            packet to flood a target device with unwanted traffic, 
            which may result in a denial-of-service condition and 
            information exposure." 
         probability = 0.0  
         CVE = "CVE-2021-38425" 
         CVSS = "CVSS:3.1/AV:N/AC:L/PR:N/UI:N/S:U/C:H/I:N/A:H"  
         BaseScore = 9.1 
         ImpactScore = 5.2 
         ExploitabilityScore = 3.9
    }
}
\end{lstlisting}

\subsection{Fault Tree Model}\label{sec:faultTreeModel}
\noindent\textbf{Summary:} The \gls{FT} model is used to capture possible faults that can endanger humans or harm the safe operation of the system.
Our proposed tool pipeline uses \glspl{FT} to generate \glspl{AFT} automatically at runtime. 
Therefore, it is required to indicate which events of a \gls{FT} can also be caused by malicious actions.
For example, in terms of the \gls{FT} in Figure~\ref{fig:exampleFaultTree}, the channel that sends messages to control the trajectory of the drone is more likely to be attacked than the rotor of a drone.

To specify events that can also be caused by malicious actions, we propose an extension of the \gls{FT} notation with attack events.
An attack event describes the event caused by a malicious action and defines conditions that an attack needs to fulfill to trigger the respective attack event.  
We use the notation of CVSS vectors to specify the conditions for an attack.
Each attack event also references either a model element from the Dataflow model or the Deployment model to specify the context of an attack.

\begin{figure}[H]
\centerline{\includegraphics[width=0.7\columnwidth]{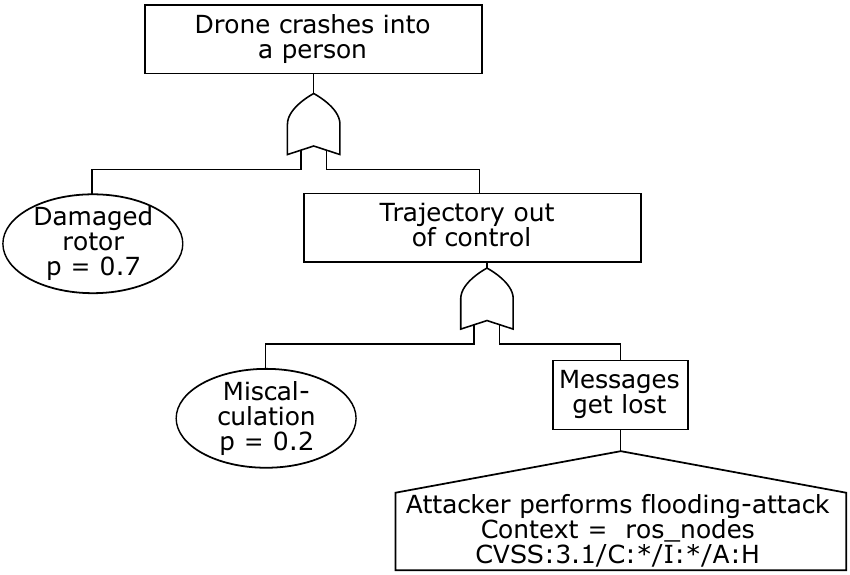}}
\caption{Extended Fault Tree including an attack step (house-shaped node). This reflects that some faults (i.e., message loss) can result from intentional attacks and not just accidental failures.}\label{fig:exampleFaultTreeExtended}
\end{figure}

Figure~\ref{fig:exampleFaultTreeExtended} shows an example of a \gls{FT} that is extended with an attack event~(house shape).
This attack event specifies that an attack that leads to loss of messages needs to target \textit{ros\_nodes}.
This is a reference to the respective (abstract) model element in the Deployment model that represents the platform encompassing all ROS nodes in the system.
Based on the definition of CVSS vectors, the attack event specifies that an attack needs to have a high impact on the availability~(A:H).
Other properties, such as confidentiality~(C) or integrity~(I) are not significant for this particular attack type.
Thus, we assign them~* meaning  ``don’t care'' in our extension.\\
\noindent\textbf{Obtained:} Manually crafted by a safety expert.\\
\lstset{style=SaftModelStyle}
\begin{lstlisting}[caption={Textual representation of the Fault Tree in Figure~\ref{fig:exampleFaultTreeExtended}.},label=faultTreeListing]
Hazard description = "Drone crashes into a person"{
  OR{
    BasicEvent description="Damaged rotor" probability=0.7,
    IntermediateEvent description="Trajectory out of control"{
      OR{
        BasicEvent description = "Miscalculation" 
         probability=0.2,
        IntermediateEvent description = "Messages get lost"{
          AttackEvent description="Attacker performs flooding-
                                   attack" 
          deploymentElement=ros_nodes 
          CVSSREQ=CVSS:3.1/C:*/I:*/A:H
        }
      }
    }
  }
}	
\end{lstlisting}

\noindent\textbf{Textual Representation:} Listing~\ref{faultTreeListing} shows an example of the textual representation we designed to denote our extended notation of \glspl{FT}.
We define for each event type a corresponding keyword~(\texttt{\color{keywordcolor}{Hazard}}, \texttt{\color{keywordcolor}{BasicEvent}}, \texttt{\color{keywordcolor}{IntermediateEvent}}, \texttt{\color{keywordcolor}{AttackEvent}}) followed by a description of the respective event and its probability.
Since attack events need to specify the preconditions an attack needs to fulfill to trigger the corresponding event, our textual representation also provides keywords to define the requirements for an attack using the notation of CVSS vectors~(\texttt{\color{keywordcolor}{CVSSREQ}}) as well as to specify the context of an attack.
We define terminal rules to enforce the correct use of the notation of CVSS vectors.
The context of an attack can either be defined as a reference to a model element in the Dataflow model, using the keyword \texttt{\color{keywordcolor}{dataflowElement}}, or as a reference to a model element in the Deployment model, using the keyword \texttt{\color{keywordcolor}{deploymentElement}}.

\subsection{Attack-Fault Tree Model}\label{sec:attackfaultTreeModel}
\noindent\textbf{Summary:} The \gls{AFT} model combines safety and security aspects. 
We use \glspl{AFT} as an intermediate representation, which we transform into \glspl{DFT} (see Section~\ref{sec:models_dft}).
The transformed \gls{AFT} is then used as input for the Storm model checker to analyze the occurrence probability of the root hazard.
We describe our approach for attaching \glspl{AT} to \glspl{FT} to create \glspl{AFT} in Section~\ref{sec:comibationOfModels}.\\
\noindent\textbf{Obtained:} Automatically generated by extending a \gls{FT} by attaching \glspl{AT} based on data from the Dataflow and Deployment models.\\
\noindent\textbf{Textual Representation:} Since the \gls{AFT} model is a combination of the \gls{AT} model and the \gls{FT} model, our textual representation is just a combination of textual representations of these two models.
Listing~\ref{attackFaultTreeListing} shows an example of our textual representation of the \gls{AFT} that is portrayed in Figure~\ref{fig:exampleAttackFaultTree}.
\lstset{style=SaftModelStyle}
\begin{lstlisting}[caption={Textual representation of the Attack-Fault Tree in Figure~\ref{fig:exampleAttackFaultTree}.},label=attackFaultTreeListing]
Hazard description = "Drone crashes into a person"{
  OR{
    BasicEvent description="Damaged rotor" probability=0.7,
    IntermediateEvent description="Trajectory out of control"{
      OR{
        BasicEvent description="Miscalculation" probability=0.2,
        IntermediateEvent description="Messages get lost"{
          IntermediateEvent description ="Perform flooding-attack"{
            AND{
              AttackStep description="Identify 
                          receiving components" probability=0.1,
              AttackStep description="Generate traffic"
                          probability=0.8 
            }		
          }
        }
      }
    }
  }
}
\end{lstlisting}

\subsection{Dynamic Fault Trees}\label{sec:models_dft}
\noindent\textbf{Summary:} \glspl{DFT} were first introduced in 1992 by Dugan et al. \cite{DFTs}. They extend \glspl{FT} by additional gates such as functional dependency (FDEP), priority AND (PAND), and simultaneous AND (SAND) to capture dynamic behavior, which in turn are reused in \glspl{AFT} (see Section~\ref{sec:backgroundfaultTrees}). A \gls{DFT} can be converted into a Markov model to calculate the overall failure rate for given failure rates of basic events.

Since \glspl{AFT} can be considered as \glspl{DFT}, which are extended by attaching \glspl{AT} with the help of PAND gates~\cite{Stoelinga4}, we follow this notion and transform \glspl{AFT} into \glspl{DFT} using an exogenous model transformation~\cite{MENS2006125}. 
Transforming \glspl{AFT} into another modeling formalism is a common analysis approach and is usually done by defining the translation of individual \gls{AFT} elements into their counterparts in the goal-formalism, as mentioned in other related work~\cite{Stoelinga2,Stoelinga1,9224539}.

We translate \glspl{AFT} to Galileo, a common language for DFTs \cite{Galileo2000} which can be understood by a plethora of tools (see Section~\ref{sec:modelchecking} for more details). The resulting DFT can be analyzed e.\,g.\ by a continuous-time Markov chain (CTMC) model checker such as Storm\footnote{https://www.stormchecker.org/}, resulting in a probability value representing the MTTF of the top node of the original \gls{FT}. 

Although the syntax is very similar between \glspl{FT}, \glspl{AFT}, and \glspl{DFT}, there are some slight differences regarding their semantics \cite{Stoelinga4}. Additionally, \glspl{DFT} themselves exist in different dialects, each with its own subtle semantic variations \cite{DFTsUncovered}. Nevertheless, we decided to use \glspl{DFT} as our input model for a model checker instead of e.\,g.\ stochastic timed automata (STA) (for a translation of \glspl{AFT} to STAs see \cite{Stoelinga1}) because of a) the straightforward translation, b) the much faster model checking, and c) we are not interested in the absolute resulting value, but in its change over time (see Section~\ref{sec:modelchecking}). \\
\noindent\textbf{Obtained:} Automatically derived from \glspl{AFT} by an exogenous model transformation.\\
\noindent\textbf{Textual Representation:} Listing~\ref{lst:dftListing} shows an excerpt of the generated DFT derived from the AFT in Listing~\ref{attackFaultTreeListing}. 
\begin{lstlisting}[caption={Shortened excerpt of the dynamic fault tree generated for our running example.},label=lst:dftListing]
toplevel Drone_crashes_into_a_person; 
Drone_crashes_into_a_person or 
    Trajectory_out_of_control Damaged_rotor;
Damaged_rotor lambda=0.7; 
Trajectory_out_of_control or 
    Miscalculation Messages_get_lost;
Miscalculation lambda=0.2; 
Messages_get_lost or 
    Poor_connection Too_many_incoming_messages;
Too_many_incoming_messages or 
    Poor_configuration_of_a_node 
    Attacker_performs_flooding_attack_on_trajectory_assignment_topic;
Attacker_performs_flooding_attack_on_trajectory_assignment_topic and 
    Determine_communication_mechanism  
    Position_in_between_the_target 
    Attacker_performs_denial_of_service_attack;
Attacker_performs_flooding_attack_on_trajectory_assignment_topic_SEQ seq 
    Determine_communication_mechanism
    Position_in_between_the_target 
    Attacker_performs_denial_of_service_attack;
...
\end{lstlisting}

\section{Running the SAFT-GT pipeline}\label{sec:SAFT-GTPipeline}
After introducing the different models, we take a closer look at the ``heart'' of our approach. We explain in detail, how the combination of \glspl{FT} and \glspl{AT} into \glspl{AFT} work and how the resulting \gls{AFT} is translated into a \gls{DFT}, the input model of a \gls{PMC} such as Storm. 

As shown by the icons in the lower right corner in Figure \ref{fig:overview}, our approach welcomes manual additions. All automatically created or derived models can be adapted or supplemented manually at any time. This ensures that existing models can be reused.

\subsection{Combination of the different models}\label{sec:comibationOfModels} 
In this section, we describe how we attach generated \glspl{AT} to manually crafted \glspl{FT} based on context information provided by the Dataflow and Deployment model.
This combination process is shown in step 5. ``Fragment Combination Model Transformation'' in Figure~\ref{fig:overview}.

In contrast to our \gls{AT} example in Figure~\ref{fig:exampleAttackTree}, the generated \glspl{AT} are at a different level of abstraction (see Listing~\ref{attackTreeListing}).
Thus, there is also a difference in the abstraction level between manually created \glspl{FT} and generated \glspl{AT}.
The events in our \gls{FT} describe --- on a rather abstract level --- the causes for a hazard without considering the concrete implementation details.
However, the generated \gls{AT} describes specific vulnerabilities of the used components.
For example, the \gls{AT} in Listing~\ref{attackTreeListing} is based on a vulnerability that can be exploited to perform a flooding attack. Nonetheless, the necessary intermediate steps to achieve this are missing if we just attach the generated \gls{AT} to the \gls{FT}.
To solve this problem, we define \gls{AFT} fragments, which describe the necessary intermediate steps to perform attacks.
These \gls{AFT} fragments are manually defined by the authors with modeling security expertise and are based on attack descriptions from the MITRE ATT\&CK framework.
The already defined \gls{AFT} fragments are part of the pipeline and can therefore be reused by all its users.
The \gls{AFT} fragments are attached to the \gls{AFT} before we attach the generated \glspl{AT} to mitigate the different abstraction levels between \glspl{FT} and generated \glspl{AT}.
To accomplish this, we divided the \gls{AFT} generation process we developed into three phases. 

In the first phase, the manually crafted \gls{FT} model, which is provided as input for the generation process, is transformed into an \gls{AFT} model.
This transformation is a simple horizontal, exogenous transformation~\cite{MENS2006125} (since the \gls{FT} represents the top of our \gls{AFT}), as illustrated in Section~\ref{sec:attackfaultTreeModel}.
It should be noted that we extend the \gls{FT} notation to include so-called Attack Events (cf. Section~\ref{sec:attackfaultTreeModel}).
These Attack Events are also added to the \gls{AFT} and form the possible attachment points for attacks in the following phases.

\begin{figure}[H]
\centerline{\includegraphics[width=0.3\columnwidth]{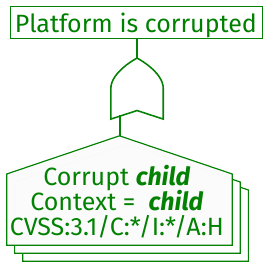}}
\caption{AFT fragment modeling the corruption of a platform by corrupting its components.}\label{fig:fragmentExample}
\end{figure}

The second phase deals with different abstraction levels between \gls{FT} and generated \glspl{AT}.
For this purpose, we have defined \gls{AFT} fragments which describe the necessary intermediate steps to perform common attacks.
This allows us to break a more abstract attack scenario, e.g., a flooding attack, into multiple atomic steps, such as  ``corrupting a platform''. Figure~\ref{fig:fragmentExample} shows an example of \gls{AFT} fragments.

For each \gls{AFT} fragment, we define conditions that are checked against the requirements defined within each Attack Event in the \gls{AFT} from phase one.
These provided conditions consist of two parts, namely the context and the attack property.
The context describes what type of model element is affected by the attack.
For example, an Attack Event must reference a model element from the Deployment model of the type \texttt{\color{keywordcolor}{Platform}}.

The second provided condition represents the property of an attack and this is defined for each \gls{AFT} fragment.
To express this property, we use the notation used by CVSS vectors.
Since the entries in CVSS vectors are on the nominal and ordinal scale, we developed a comparison mechanism to check that the provided conditions of an AFT fragment fulfill the requirements of an Attack Event.
Accordingly, we compare values such as Attack Vector~(AV), User Interaction~(UI), or Remediation Level~(RL) that lie on a nominal scale for equality.
In regard to the values that lie on an ordinal scale, we compare them with ``greater-than-equal''.
This means, for example, that if an Attack Event defines that Confidentiality Requirement~(CR) should have the value Medium~(M), the provided conditions of the AFT fragment must define that CR is Medium~(M) or High~(H).
For the \gls{AFT} fragment in Figure~\ref{fig:fragmentExample}, we defined that Availability~(A) must be High~(H).
This means that this \gls{AFT} fragment can only replace Attack Events in the \gls{AFT} that reference the respective context and define~A=H.
If this fragment can be applied, it creates a new Attack Event for each component that is part of the previously referenced platform, describing that it is being corrupted (represented in Figure~\ref{fig:fragmentExample} by the overlapping house shapes and \textbf{\textit{child}} as a placeholder).

Regarding our running example, this means that after the \gls{FT} in Figure~\ref{fig:exampleFaultTreeExtended} was transformed into an \gls{AFT} in the first phase, the Attack Event ``Attacker performs flooding-attack'' remains.
In the second phase, our \gls{AFT} generation approach checks whether there are \gls{AFT} fragments whose preconditions fulfill the requirements defined in the Attack Event.
This is the case with the \gls{AFT} fragment in Figure~\ref{fig:fragmentExample}.
Thus, the Attack Event is replaced with the fragment, and for each component that is part of \textit{rose\_nodes}, a new Attack Event is generated and connected to the intermediate event ``Platform is corrupted'' by an OR gate.

The third phase deals with attaching the generated \glspl{AT} to the \gls{AFT} resulting from the previous two phases.
For this purpose, the referenced context is considered for each Attack Event. 
More precisely, each model element from the Deployment model referenced in the \gls{AFT} is recursively scanned to determine whether it has a vulnerability or consists of/uses a component with a vulnerability.
If a Dataflow element is referenced by an Attack Event, the corresponding model element in the Deployment model is identified before the recursive analysis. 
Consequently, when a system component with a vulnerability has been identified, we re-use the mechanism described in phase two to ensure that the requirements described using the CVSS notation are also served by the discovered vulnerability.
This leads to the corresponding Attack Event in the \gls{AFT} being replaced by an \gls{AT}.
If an Attack Event can be replaced by several \glspl{AT}, we combine them using OR gates to attach them to the \gls{AFT}. 

\begin{figure}[tb]
\centerline{\includegraphics[width=0.8\columnwidth]{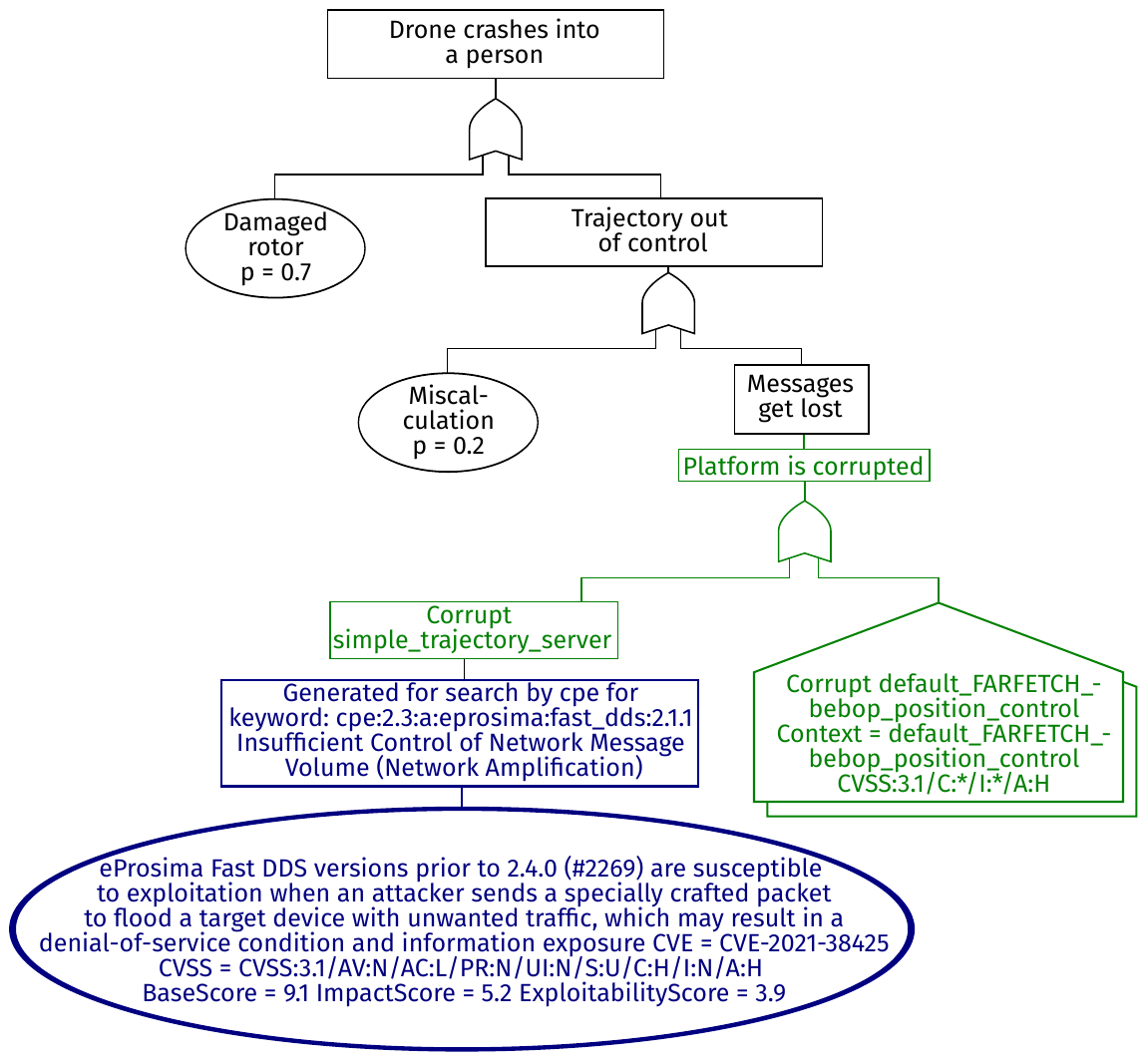}}
\caption{Simplified excerpt from the AFT generated by our approach.}\label{fig:genAFTExample}
\end{figure}

In our example, the generated \gls{AT} in Listing~\ref{attackTreeListing} fulfills the conditions of the attack event that was generated as part of the \gls{AFT} fragment in Figure~\ref{fig:fragmentExample} for the Dataflow model element \textit{simple\_trajectory\_server}.
As a result, this Attack Event is converted into an intermediate event and the generated \gls{AT} is attached to it.
Thus, we obtain the \gls{AFT} shown in Figure~\ref{fig:genAFTExample}.

\subsection{Risk Calculation}\label{sec:modelchecking}

The resulting \gls{AFT} has to be analyzed via a \gls{PMC} returning the MTTF of the current system configuration. Thus, each \gls{AFT}  has to be transformed into an appropriate input language of a model checker. Stöelinga et al. present in \cite{Stoelinga1} a translation of \glspl{AFT} to stochastic timed automata~(STA)~\cite{STA2011}. For model checking of this formalism, the UPPAAL Stochastic Model Checker~(SMC)~\cite{uppaalSMCtutorial2015} can be used. Unfortunately, UPPAAL SMC only allows simulation and no closed-form solution. This means that the results depend on the number of simulated cycles and can vary due to randomness. For reliable results, 100,000 or more cycles are necessary and require corresponding significant computation times.  

Instead, we decided to extend the toolchain with a closed-form solution provided by a continuous-time Markov chain~(CTMC) model checker such as Storm. Although there are a couple of similarities and differences in the sense of syntax and semantics between \glspl{FT} and \glspl{AT}, we follow the conclusion from~\cite{Stoelinga4} that ``\glspl{AFT} trade scalability for versatility, by merging \glspl{DFT} (with all its dynamic gates) with \glspl{AT} plus PAND gates.'' \glspl{DFT} in Storm already supports exponential probability distribution as well as OR and AND gates. PAND semantics can be simulated by combining an AND gate with a SEQ gate with the same children.

Regarding the probability of a successful attack, we assume an exponential distribution that reaches the given EPSS value after 30 days \cite{epss2021}. Thus, the obtained EPSS scores of the Basic Attack Events are translated into a rate $\lambda$ according to the following formula:
\begin{equation}
    \lambda = -\frac{\ln(1-\text{epss})}{30\cdot24\cdot60\cdot60s}
\end{equation}
For specific attacks modeled in manually created \glspl{AT}, this value has to be provided by the modeler. 

The resulting \glspl{DFT} are then passed to the Storm model checker to calculate the MTTF (according to Equation 1 in seconds) of the top event of each original \gls{FT}. An MTTF indicates how long it takes in average until the top event fails. Estimation of the various probabilities is an extremely difficult task, even when the EPSS scores are taken into account. Due to this uncertainty, we suggest not considering the absolute MTTF but its change over time to get insights, if the overall risk (of this particular) hazard in- or decreased, e.\,g.\ due to newly discovered CVEs\footnote{We acknowledge that this transformation assumes a monotonic increase in exploit likelihood, which may not hold for all vulnerabilities. Thus, the resulting rates should be interpreted as approximations.}. If the value decreases, a more secure and thus also more reliable state has been reached. Finally, the resulting MTTF is fed back to the MAPE-K system in order to integrate the new information into new adaptations, if necessary. 

\noindent\textbf{Interpretation of probabilities.} The estimation of attack and failure probabilities remains challenging computationally and conceptually. Prior works do not provide consensus on how to meaningfully combine stochastic failures with deliberate attacks. 
Our interpretation follows the line of work by Nai Fovino et al.~\cite{fovino2009integrating} and Kumar and Stoelinga~\cite{Stoelinga1}, who demonstrated that treating attack events as probabilistic basic events within AFTs is methodologically sound when the probabilities are grounded in empirical exploit likelihoods (e.g., EPSS/CVSS). 
Accordingly, the MTTF values in our analysis should not be interpreted as absolute predictions of joint safety–security risk\footnote{For instance, cybersecurity outcomes can be strongly influenced by organizational context and human factors, which can introduce variability that is not reflected in purely probabilistic or data-driven models \cite{pfister2025department}.}. On contrary, they should be taken as relative indicators used to compare configurations and runtime adaptation decisions within the same system context.

\subsection{Closing the Loop}
\label{sec:closingTheLoop}
We integrate the toolchain into the MAPE-K loop in a way that allows us to trigger each step separately. Together with the possibility to reuse already created artifacts, this allows for an incremental approach depending on the changes that occurred in the overall system. In case of a reconfiguration, the dataflow within the analyzed system might have changed and thus, a complete rerun is necessary. This means that the dependencies might have changed with newly activated nodes, which can include new components and libraries. For this reason, it is necessary to identify new vulnerabilities that would affect these new nodes and these need to be discovered. In the case of a system update (or migration of particular ROS2 nodes to another underlying system), it is enough to update the dependencies of the deployed components and thus, the first two steps can be skipped. Finally, in cases when the public CVE database is updated --- for example, once a day --- a new \gls{AT} generation can be triggered after this update. In regard to the EPSS values, the EPSS database is updated only once a year\footnote{\url{https://www.first.org/epss/model}}.

Since the execution of the different parts of the pipeline might take a certain amount of time (see Section~\ref{sec:experiments}), the ROS2 service call just validates the input parameters and returns immediately. The resulting MTTF value is sent asynchronously via a separate ROS topic. This enables the use of multiple receivers and can also be used to signal to the MAPE-K system that a triggered toolchain run has finished.

\section{Evaluation}
\label{sec:evaluation}

To determine the overall approach's perceived relevance and applicability, we conducted a study where subject-matter experts of varying acumen were asked to provide their opinions on the developed approach. In this regard, a survey and a comprehensive discussion with the participants were conducted. Moreover, we provide an experimental evaluation of the proposed approach. This was also done because we did not provide any evaluation of the presented approach in our previous works due to the large scope and the considerable effort that was required.

\subsection{User Study Design}\label{sec:study_design}
The study was designed as a part of a two-hour workshop, which consisted of two surveys, three presentations (\textit{SAFT-GT Overview and Agenda}, \textit{Input Models and Model Generation}, and \textit{AFT Model Combination and Analysis}), and three discussions (\textit{\gls{SAS}: Experiences and Challenges}, \textit{Safety and Security Models and Modeling}, and \textit{SAFT Toolchain and Model Combination}). Each presentation was followed by a discussion, which targeted that particular presentation. The workshop was recorded and was agreed to be recorded by all the participants. All participants were formally informed and provided written consent. This included consent for data collection, processing, anonymization and eventual publication of their anonymized responses. Consent also explicitly covered the linkage between discussion data and surveys and participants were assured of their right to withdraw at any time. In order to promote transparency and reproducibility of the user study, we release transcribed discussions, linked survey questions and consent form in our open-source repository (see Section \ref{sec:introduction}) to support independent verification and replication efforts.
\\

\noindent\textbf{Participants.} As a part of the recruitment process, we reached out to multiple experts, each having expertise in one or more fields that tackle the context of our developed approach. They were asked if they were willing to partake in a workshop by providing them with a short description of our research project. We conducted all communication via email. In total, seven experts gave a positive response and they were provided with a link to a survey\footnote{This survey is independent of the two that were conducted during the workshop.} in which we anonymously collected some background information. This was done one week before the workshop. Each of the seven experts was from the research domain. In addition, some of them also had additional backgrounds as research and development consultants in the industry domain. They all had at least 3-5 years of experience in safety modeling and engineering of cyber-physical systems (three of them had 10+ years of experience). In regards to security modeling, the experience ranged from 1-9 years. Moreover, all the experts had knowledge and experience in the following domains: domain-specific languages, self-adaptive systems, and \glspl{FT}. However, the participants had slightly less expertise concerning security topics and domains utilized in our proposed extended approach. This relates to CVEs, CVSS, CWEs, EPSS, \glspl{AT}, and \glspl{AFT}. Nonetheless, this is the case with only one or two participants. Thus, it can be concluded that the safety background of experts was on the stronger side compared to their security background.\\

\noindent\textbf{Survey.} Two surveys were conducted during the workshop. One survey focused on the applicability of the proposed extended approach, while the other one targeted its relevance. Every question in the surveys was answered by all the participants of the workshop. The answers to each of the questions were provided in accordance with the Likert scale: \textit{very (applicable/relevant)}, \textit{(applicable/relevant)}, \textit{neutral}, \textit{not (applicable/relevant)}, and \textit{not (applicable/relevant) at all} for applicability and relevance respectively. We hosted surveys on unpublished websites, and we never asked for participants’ sensitive or personal information.\\

\noindent\textbf{Expert opinions.} Expert opinions were collected at three different intervals during the workshop. This was done after each presentation. The discussions took 10, 20, and 20 minutes, respectively. The first discussion targeted various challenges and experiences with \gls{SAS}. This provided us with insights regarding how experienced the experts were in this particular topic. In the second discussion, the relationship between safety and security models, as well as their combination, was addressed. Finally, the last discussion examined the approach that we extended, including the full toolchain and model combinations. During this process, we adhered to the relevant empirical standards for qualitative studies as outlined in the SIGSOFT Empirical Standards\footnote{\url{https://www2.sigsoft.org/EmpiricalStandards/docs/standards?standard=QualitativeSurveys}}.

\subsection{Results}

\begin{figure*}[htb]
\centering
\sidesubfloat[]{
    \includegraphics[width=0.99\textwidth]{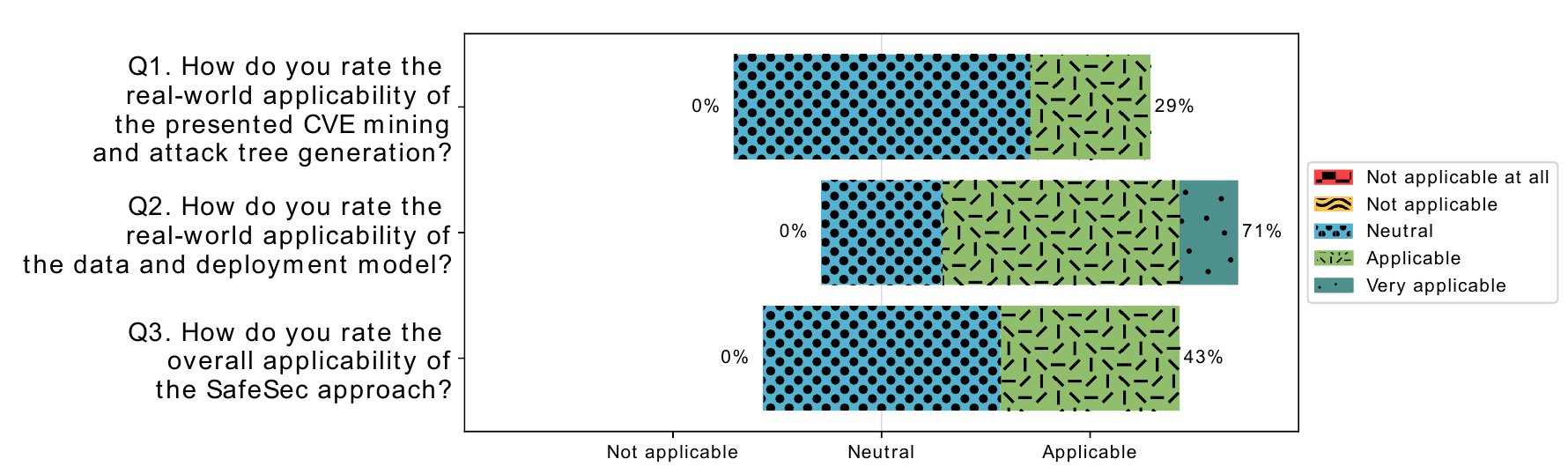}
    \label{fig:first}
}\\
\sidesubfloat[]{
    \includegraphics[width=0.99\textwidth]{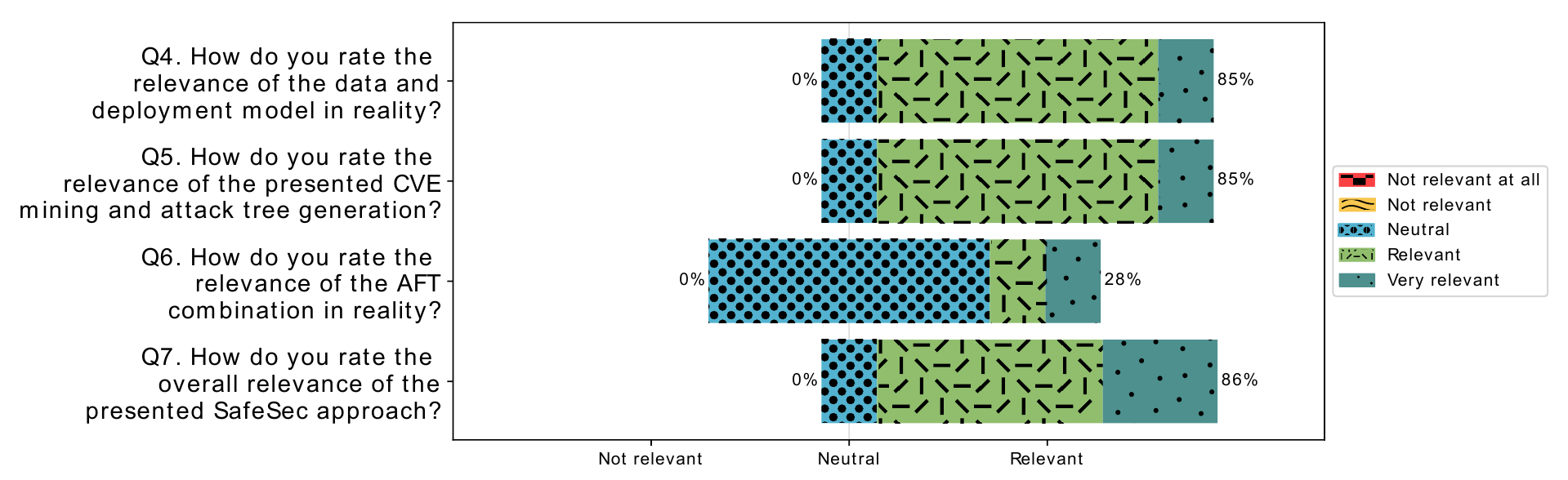}
    \label{fig:second}
}
\caption{Workshop Survey Response Results}
\label{fig:figures}
\end{figure*}

\noindent\textbf{Survey.} Each of the seven participants provided answers to all of the questions from both surveys. Figures~\ref{fig:first} and \ref{fig:second} present the results regarding the applicability\footnote{The applicability survey has only three questions because it was executed before the AFT combination topic was covered in the workshop.} and relevance of the developed approach respectively\footnote{The percentage values were rounded so they may not sum up to 100\%.}. The findings revealed that in Q1, 29\% of experts rated the approach as applicable, while 71\% were neutral. In Q2, 14\% found the approach very applicable, 57\% rated it as applicable, and 29\% were neutral. Furthermore, the overall applicability of the approach received ratings of 14\% applicable and 86\% neutral in Q3. 

In terms of relevance, 14\% found the data and deployment model very relevant, 71\% rated it as relevant, and 15\% were neutral in Q4. For the relevance of the presented CVE mining and \gls{AT} generation approach, 14\% found it very relevant, 71\% rated it as relevant, and 15\% were neutral in Q5. Moreover, regarding the relevance of the AFT combination in reality, 14\% found it very relevant, 14\% rated it as relevant, and 72\% were neutral in Q6. Lastly, the relevance of the overall approach was rated as 28\% very relevant, 58\% relevant, and 14\% neutral in Q7. These responses provide valuable insights into the perceived applicability and relevance of the approach, indicating areas for further consideration and potential improvement, which will be covered in Section \ref{sec:discussion}.\\

\floatsetup[figure]{style=plain,subcapbesideposition=top}

\noindent\textbf{Expert opinions.} After carrying out the discussions, we qualitatively analyzed the recording of the workshops (we cannot share the recording due to NDA). The aspects covered can be organized into the following five topics:\\

\begin{itemize}
    \item \textbf{Design vs. Runtime Approach}: The question of ``\textit{why the approach was not implemented at design time}'' was raised by one of the experts. However, the other expert stated: ``\textit{There will be some unexpected conditions. There exist many states that cannot be foreseen at design time, which can be recorded quite well by monitoring the system runtime. These states of the system can actually be recorded at runtime, including any unexpected conditions. There is also the case that not everything is known at design time. Perhaps new things will be added.}'' According to the aforementioned statements, the design time approach was found to be challenging due to unexpected conditions and the necessity for all components to be known in advance. Moreover, the possibility of adding new components to the system that cannot be predicted at design time was acknowledged. This also includes the possible space exploration in regards to the system and its environment, which may lead to a state space explosion. Therefore, addressing these concerns from the design time perspective may be infeasible and experts agreed that the runtime perspective was appropriate, considering the dynamic nature of system components and conditions.\\
    \item \textbf{Closed vs. Open System}: The expert stated: ``\textit{Simply because there are a lot of systems that include networking, you have an incredibly large attack surface because of the networking itself. But once the system is closed, at least physically, and there is no flow of information to the outside and only the sensors are the attack surface.}'' Thus, the proposed approach was deemed suitable for open systems, as it accounts for the networking aspect, resulting in a larger attack surface. In contrast, for closed systems where the networking aspect is absent, the focus would need to be shifted to attacks occurring over sensors.\\
    \item \textbf{Trusting Experts vs. Probabilities}: Another expert noted the following: ``\textit{There are many successful attacks that occur in everyday life that are not recorded, which require significant expert knowledge in order to be addressed. Do we then want to rely on these calculated figures? I would be skeptical at first.}'' Based on the aforementioned statement, it was evident that the implementation raised concerns regarding trust in automated evaluations versus manual expert assessments. Despite the prevailing tendency to trust manual evaluations more, the increasing complexity of systems necessitates the adoption of automated approaches. The implementation supports the need for continuous system checks due to the inherent possibility of unforeseen events, reinforcing the importance of automated evaluations. Nevertheless, the proposed approach includes the possibility of experts adding additional states manually, such as zero-day attacks.\\
    \item \textbf{Modularity}: During the workshop, the following was mentioned: ``\textit{Several data models that reference different components can be simply plugged together, which at the end form a complete system. For example, even different competitors can share model fragments. However, then it is not very clear at which point they should probably share something.}'' Thus, the approach's ability to generate \glspl{AFT} for specific parts or components of a system was emphasized. This modularity facilitates the sharing of \gls{AFT} fragments among industry competitors for specific system parts, reducing the cost of evaluating larger systems by allowing fragments to be plugged into other \glspl{AFT}, which makes the evaluation of larger systems less costly.\\
    \item \textbf{Variety of Security Attacks}: One of the experts mentioned: ``\textit{But the first that came to my mind was, well, if you look at the functional safety of systems and if you can significantly change the system in operation... This refers specifically to \glspl{SAS}, then you open up a whole additional world of possibilities and attacks that can influence the system's behavior considerably.}'' Another expert said: ``\textit{If there is an attack that is not publicly recorded and I would like to add this attack manually, this does not bring too many advantages. This is because a lot of things need to be done manually by creating or updating separate models. However, these could theoretically be used again and inserted into this automated setup here. I see that models are theoretically integrable in this modeling approach and that automation depends a lot on the data flow model. Also, if, for example, in the hardware I have somehow a hardware component such as an electric motor and there are magnetic beams that lead to bit flips that target the RAM memory... Then that is not about the data flow model and requires additional consideration, which requires the further extension of the proposed approach.}'' Based on the aforementioned statements, it can be noted that the experts mentioned it is necessary to consider attacks when the system adapts to new configurations, which highlights the potential of targeting reconfiguration mechanisms by attackers. In addition, attacks targeting adaptation mechanisms need to be addressed. While the approach relies on public security databases, it allows for the manual addition of other types of attacks and emphasizes the necessity to consider mechanical attacks. In cases involving mechanical attacks (such as using beams to create bit flips), traditional models like the data-flow model may not be as relevant, necessitating a broader consideration of attack vectors. 
\end{itemize}

\vspace{3.5mm}

\textbox{
\textbf{Disclaimer:} The statements above stem from our own interpretation of the (sparse and unstructured) discussions. However, they reflect the opinion of experts involved in the workshop. 
}

\subsection{Performance} 
\label{sec:experiments}
The following opinion was stated by an expert: ``\textit{Industrial experts, who are in this particular field of expertise, would need some assurances regarding if the developed modeling approach and generation of specific models is too expensive. In my previous experiences, which specifically relate to hazard and risk analysis, it is important to know and measure the costs of running such implementations or toolchain.}'' We agree with this statement, especially because our previous work~\cite{groner2023model} does not contain such evaluation.

In general, it is not easy to provide a holistic experimental evaluation of this approach, as it strongly depends on many parameters: the size of the models, the amount and size of \glspl{FT}, manually added \glspl{AT}, vulnerabilities found in a system, to name but a few. Nevertheless, we set up the ROS2 quadcopter example mentioned above in our quadcopter lab and measured the elapsed time for each of the steps in the integrated toolchain. This was done to determine which steps are expensive and if the toolchain could scale.  

In our setup, the entire ROS2 part (managed system, MAPE-K loop) ran on a different machine than the toolchain (both i7-3770 @ 3.40GHz, 16GB RAM). The dataflow model of our ROS-based example consists of $21$ nodes that are connected via 43 ROS topics. We created 3 \glspl{FT} manually with 13 nodes/4 gates (\textit{battery}), 11 nodes/6 gates (\textit{spying}), and 33 nodes/9 gates (\textit{injure}). The deployment model generator extracted 265 dependencies on the machine running both the managed system and the MAPE-K loop.

Our pipeline mined a local copy of the NVD database with around $272,400$ entries and identified $77$ CVEs affecting system parts connected to causing the hazard modeled by our \gls{FT}.
Based on these CVEs, our pipeline generated 5 \glspl{AT} by matching CPEs ($2$\,$\times$\,$1$ CVE, $1$\,$\times$\,$8$ CVEs, $1$\,$\times$\,$17$ CVEs, $1$\,$\times$\,$18$ CVEs). Another 6 \glspl{AT} were generated by full text search in the CVE's description, which is more likely to lead to false positives ($2$\,$\times$\,$1$ CVE, $2$\,$\times$\,$2$ CVEs, $1$\,$\times$\,$3$ CVEs, $1$\,$\times$\,$23$ CVEs).

The \glspl{AFT} combined from the \glspl{FT} and the amount of \glspl{AT} has three different sizes: The smallest (battery) has $38$ nodes and $14$ gates, the mid-sized (spying) has $926$ nodes and $461$ gates. Finally, the largest \gls{AFT} (injure) has $1845$ nodes and $814$ gates. The \glspl{DFT} generated from the resulting \glspl{AFT} and served as input for the Storm model checker have between $26$ nodes and $14$ gates up to $990$ nodes with $918$ gates. This difference arises because some nodes can be combined and some additional intermediate nodes have to be added during the conversion.

\begin{table}[htbp]
\caption{Results of Running Example Model Generation and Evaluation Experiment}
\label{tab:results}
\begin{center}
\begin{tabular}{|S[table-format=3.3]|S[table-format=3.3]|S[table-format=3.3]|S[table-format=3.3]|}
\hline
& \multicolumn{3}{c|}{\textbf{Model Generation Time}}  \\
\cline{2-4} 
\multicolumn{1}{|c|}{\textbf{[s]}} & \multicolumn{1}{c|}{\textbf{Datafl.}} & \multicolumn{1}{c|}{\textbf{Deploy.}}& \multicolumn{1}{c|}{\textbf{AT}} \\
\hline
\multicolumn{1}{|c|}{avg} & 4.979 & 273.925 & 552.596 \\\hline
\multicolumn{1}{|c|}{sd} & 1.022 & 0.743 & 1.457 \\\hline
\end{tabular}

\begin{tabular}{|S[table-format=3.3]|S[table-format=3.3]|S[table-format=3.3]|S[table-format=3.3]|S[table-format=3.3]|S[table-format=3.3]|S[table-format=3.3]|}
\hline
& \multicolumn{6}{c|}{\textbf{Model Generation/Analysis Time}} \\
\cline{2-7} 
& \multicolumn{2}{|c|}{battery} & \multicolumn{2}{|c|}{spying} & \multicolumn{2}{|c|}{injure} \\
\cline{2-7}
\multicolumn{1}{|c|}{\textbf{[s]}} & \multicolumn{1}{c|}{\textbf{AFT}} & \multicolumn{1}{c|}{\textbf{DFT+MC}}& \multicolumn{1}{c|}{\textbf{AFT}} & \multicolumn{1}{c|}{\textbf{DFT+MC}} & \multicolumn{1}{c|}{\textbf{AFT}} & \multicolumn{1}{c|}{\textbf{DFT+MC}} \\
\hline
\multicolumn{1}{|c|}{avg} & 10.168 & 3.124 & 17.619 & 3.042 & 21.708 & 3.650\\\hline
\multicolumn{1}{|c|}{sd} & 1.075 & 0.270 & 1.906 & 0.088 & 0.936 & 0.127\\\hline
\end{tabular}
\end{center}
\end{table}

Table~\ref{tab:results} shows the average result of 100 runs for each step. We focus on the execution time of the toolchain and not on the duration of the reconfiguration process. This is due to the fact that the duration of the reconfiguration of the system strictly depends on the implementation of the reconfiguration mechanism, which is out of the scope of this paper. The entire toolchain can be executed in about 15 minutes ($\sim$860s), wherein the most time is needed for the full text search of the database ($\sim$10min) followed by the extraction of the package dependencies ($\sim$5min). Since repeating these steps is only necessary if the system changes (e.g., by removing or adding ROS nodes, updating the underlying system (and thus the dependencies), or adding new CVEs), we did not repeat these steps 100 times for each \gls{FT}. In contrast, the remaining steps (AFT generation and model checking (MC)) depend on the \gls{FT} and were therefore executed separately for each \gls{FT}.


In the following, we will discuss the factors influencing the size of the generated \glspl{AFT}, as these are used as input for the model checker and can therefore have a significant influence on performance.
The size of an \gls{AFT} depends on several factors.
Its size has a limited relationship with the system size and is independent of the number of hazards considered.
If several hazards are considered, this results in a forest of \glspl{FT} that leads to a forest of \glspl{AFT}.
These \glspl{AFT} must then each be examined with the aid of a model checker, whereby it can be assumed that not all modeled hazards are relevant in all possible system configurations and every environment.
The analysis of a forest of \glspl{AFT}, for example, by optimizing parallelization, is out of scope for the current state of work but should be considered as a future research direction.
As mentioned above, the system size has a secondary role, since the pure number of CVEs in a system does not influence the size of an \gls{AFT}.
As described in Section~\ref{sec:comibationOfModels}, \gls{AFT} fragments and \glspl{AT} are attached based on the preconditions defined as Attack Events in our proposed extension of \glspl{FT}~(cf. Section~\ref{sec:faultTreeModel}).
This means that the number of Attack Events and their fulfillment influences the size of an \gls{AFT}.

It should be noted that an Attack Event can be replaced by several \gls{AFT} fragments linked by an OR gate.
However, it is never possible to use all currently defined \gls{AFT} fragments, as some of them are mutually exclusive.
For example, no component can be a channel and a component, so the context defined in the Attack Event already limits the number of possible \gls{AFT} fragments.
As can be seen in Figure~\ref{fig:fragmentExample}, \gls{AFT} fragments in fact also define Attack Events.
A distinction can be made as to whether the appending of an \gls{AFT} fragment leads to several new Attack Events, as in our running example~(cf. Section~\ref{sec:comibationOfModels}).
Usually, \gls{AFT} fragments only define one or two Attack Events independently of the system.
For example, our \gls{AFT} fragment describing an adversary in the middle attack defines an Attack Event that describes the determination of the communications protocol used.

The number of attached \glspl{AT} depends on whether Attack Events reference components as a context that are related to the CVEs contained in the system.
Since we perform an elaborate analysis of the textual description of CVEs, our generated \glspl{AT} consist of some more nodes and thus have influence on the size of the generated \gls{AFT}. Although this additional full text search may lead to more false positives, we ignore this circumstance in this case study to demonstrate that our tool chain can also handle (rather unrealistic) large models.

Finally, we acknowledge that model checking offers a potential performance disadvantage.
However, our performance measurements show that the biggest performance bottleneck in our pipeline is currently the extraction of dependencies (resp., the full text database search).
Since this is part of our implementation, it makes sense to focus on optimizing this particular aspect first before focusing on the used third-party components.


\section{Future Research and Limitations} 
\label{sec:discussion}

In this section, we discuss future research directions and limitations of the proposed approach.

\noindent\textbf{Future research directions.} Considering the discussion with experts, future research should consider several directions. The following key points provide a roadmap for future research directions:

\begin{itemize}
    \item \textbf{Automation as the Key Focus}: Emphasize a strong focus on automating as much of the approach as possible, including specific models that are part of the approach. This is in stark contrast to manual investigation of security incidents in self-adaptive systems, particularly those leading to safety and reliability issues. Automation not only enhances efficiency but also ensures accuracy, thereby mitigating the risks associated with manual investigation. The expert mentioned the following: ``\textit{I am really happy when people at least think about the fact that something can go wrong, and that is what I actually like very much about this approach. What also helps to a high degree is the automation, i.e., recognizing things as soon as possible and even if they are not perfect, but at least automatically recognizing them is a big plus compared to manually filling many forms and tables, and then you have to do all the calculations.}''\\
    \vspace{-2.5mm}
    \item \textbf{Behavioral Models for System States}: Focus on developing behavioral models to capture different system states and states of the monitored environment. This can lead to the identification of state-specific \glspl{AFT}, thereby bolstering the system's resilience against potential threats. Behavioral models can provide a deeper understanding of system states, enabling proactive measures to be taken in response to potential security threats, which was mentioned by an expert: ``\textit{You may also consider the use of behavioral models and these usually closely relate to \glspl{SAS} because the system depends a lot on the state of the environment and specific conditions. This is especially significant for Fault Trees.}''\\ In a similar direction, future work could also explore the integration of attacker behavioral profiles into the analysis. While the current work focuses on observable vulnerabilities and structural system properties to achieve automation and scalability, attacker profiles (e.g., motivations, capabilities and behavior patterns) could potentially provide a more targeted security assessments. However, such modeling would require well-defined abstractions (e.g., attacker categories or skill levels) and the ability to translate these into probabilistic model parameters. Frameworks such as MITRE ATT\&CK could offer a foundation for including these aspects in a structured and extensible way. However, due to their variability and context dependency, attacker profiles are challenging to formalize and thus remain an open research direction.
    Additionally, the integration of a formal specification of the behavior of each node would allow a deeper analysis of the models. For example, one could define formal properties (e.g., in temporal logic) and use the probabilistic model checker to prove that these apply to the system.
    \vspace{-2.5mm}
    \item \textbf{Consideration of Non-Conventional Attacks}: Extend the scope of research to consider other types of attacks that may not be readily available in public databases. The expert stated: ``\textit{If you do not have any components in your system that are related to specific CVEs yet, you do not generate any Attack Trees, or do you? Systems are becoming more complex and thus offer many possibilities for adversaries and not all vulnerabilities that they exploit are reported.}'' This means that mechanical attacks and those targeting self-adaptation mechanisms also need to be considered in the future. However, these attacks (specifically mechanical attacks) necessitate novel approaches to security assessment and mitigation that would benefit the system's reliability. In addition, zero-day attacks \cite{bilge2012before} can be considered an unforeseen menace due to the fact that they exploit previously unknown vulnerabilities, for which no specific defense was prepared or implemented. Nonetheless, these represent a serious threat not only to self-adaptive systems but to any other system too. A possible direction that could tackle this issue is a utilization of NLP (natural language processing) techniques and Large Language Models (LLMs) to conduct an analysis of social networks and underground forums~\cite{arazzi2023nlp}~---~a place where adversaries communicate freely. These commonly include hints as well as detailed information on planned and ongoing attacks, including tactics and techniques. By considering a wider range of potential attacks, the security measures can be more comprehensive and adaptive, ensuring robust protection against diverse threats.\\
    
    \vspace{-2.5mm}
    \item \textbf{Development at Design Time\footnote{Development at design time needs to be complemented with the development at runtime.}}: Shift the focus towards developing the safety and security approach at design time, with an emphasis on predicting possible threats and fortifying the system's safety and security from the outset. The following was mentioned in the workshop: ``\textit{There are actually things that can be planned in advance or perhaps even have to be planned for the design side. However, you would theoretically have to consider all the faults for every configuration and every transition, so to speak. In case the adaptation is implemented in such a way that this configuration space is not even known for the designs, this becomes extremely difficult, but the design-time development can serve as a starting point.}'' Thus, it can be said that by addressing security concerns at design time, significant resources can be saved at runtime, especially in scenarios where the modeling approach becomes expensive due to the system's scale and complexity. However, it is clear that all the security and safety-related issues cannot be simply resolved at the design time due to the unpredictable nature of CPSs~---~and specifically in the case of \glspl{SAS}. Optimally, a combination of both the design and runtime approaches, such as the one described by Fournaris et al. \cite{fournaris2019design}, should be utilized. As a result, this would lead to a better overall reliability of a system.
\end{itemize}

\noindent\textbf{Limitations.} The limitation of the presented toolset is that the requests are sent to the remote computer via SSH. This could probably be optimized by combining several requests and not opening a new SSH connection for each request to the remote computer. In addition, during the performance evaluation, it can be seen that the standard deviation in the generation of \glspl{AT} is relatively high. 
This can be explained as follows: When searching for CVEs, CPEs are used in the first instance and, if no CPEs can be determined for a component, a full-text search is performed on the description of the CVE database on the other. 
The toolset uses a REST API from NIST for the CPE search, as this automatically takes the version number ranges into account. The response time of this network query varies considerably depending on the load on the server. In contrast, the full-text search on a local copy of the database is considerably faster. In our example, this only accounts for around 3s of the 62s. Obviously, one could also optimize this by using the local database for the CPE search and re-implementing the check of the matching version numbers of the packages. 

We provide a detailed description of our generation approach that relies on these models in Section~\ref{sec:comibationOfModels}. Moreover, while the toolset integrates standard libraries and public data sources (e.g., ROS2 introspection tools, CVE/CPE databases), all core components responsible for model generation, transformation and analysis are self-implemented. These external resources serve only as inputs or utilities. Thus, their potential errors have limited impact on the correctness of the toolchain’s internal logic.

In regard to the user study, the recruitment of experts who possess experience in various domains that are part of our approach proved very challenging due to their scarcity. Thus, we do not attempt to generalize the findings -- due to the small number of participants -- which prevents us from deriving quantitatively grounded conclusions. This is due to their scarcity, which is common in security-related user studies \cite{braun2023understanding} in which the participant size is rarely greater than 20. Moreover, the experts' background has a stronger safety focus compared to the security focus. As a result, it is possible that some security-related aspects and conclusions were overlooked. Finally, the experts could only grasp as much of our approach as it was presented to them during the three presentations at the workshop. This means they did not test the approach in real-world or practical settings. Hence, it is expected that the survey answers they provided are strictly related to the presented content. Nonetheless, according to the discussions conducted, it was safe to assume that they grasped the presented approach thoroughly.

\noindent\textbf{Limitations of combining safety and security probabilities.} The quantitative integration of safety-related failure probabilities and security-related attack probabilities remains debated in the safety and security-engineering communities. 
While AFTs provide a unified framework, their semantic foundations depend on assumptions that are not ``universally'' accepted. 
Safety failures can be considered random and stochastic, while security attacks are intentional and adaptive~\cite{kriaa_pietre-cambacedes_bouissou_halgand_2015,bloomfield2013security}. 
Several prior works argue that no single probabilistic metric can fully capture both aspects~\cite{kriaa_pietre-cambacedes_bouissou_halgand_2015,bloomfield2013security}.

In past, literature has demonstrated that a probabilistic treatment of attack events within fault-tree formalisms is feasible under clear assumptions. Nai Fovino et al.~\cite{fovino2009integrating} explicitly concluded that \emph{``it is possible to calculate the probability associated with the attack-tree goal exactly in the same way as it is done for the fault trees''}. 
Kumar and Stoelinga~\cite{Stoelinga1} later provided quantitative semantics for AFTs and showed their applicability for joint safety–security analyses. 
Accordingly, we interpret the probability of a fault event as the likelihood that a component fails, and the probability of an attack event as the likelihood that a known vulnerability is successfully exploited in the wild. We treat both as stochastic basic events within the same formalism.

To maintain interpretability, we deliberately restrict our model semantics to a subset of well-understood logical gates (AND, OR, PAND, and SAND). We exclude dynamic or repair gates such as FDEP or SPARE, whose cross-domain semantics remain debated in the community. 
Logical gates define combinations of events based purely on logical or temporal relationships (conjunction, disjunction, ordered conjunction, and sequential conjunction), which are considered interpretable in both cases when the input events represent random failures and attacks. Limiting the model to the four logical gates ensures semantic transparency, consistency across domains, and computational tractability when transforming AFTs to DFTs for model checking.

Our quantitative assessment does not serve as an absolute predictor of joint safety–security risk. On contrary, it should be considered as a relative indicator for comparing designs or runtime configurations of cyber-physical and self-adaptive systems. 
Nonetheless, a universally accepted probabilistic framework for combined safety–security aspects is still an open research challenge.

\noindent\textbf{Limitations of probabilistic modeling based on EPSS.} Our approach approximates the probability of successful exploitation using an exponential distribution derived from EPSS scores. This directly implies a monotonically increasing hazard rate. However, empirical EPSS trajectories are not always monotonic and may fluctuate due to factors such as patch availability or changing attacker incentives. Consequently, this assumption introduces approximation errors that can affect the accuracy of the resulting MTTF values. This is particularly the case for vulnerabilities with highly dynamic exploitation likelihoods.

To address this, MTTF is not interpreted as an absolute prediction but as a relative indicator for comparing system configurations and guiding adaptation decisions. This reduces sensitivity to modeling inaccuracies by focusing on trends rather than exact probabilities. For cases in which EPSS behavior deviates significantly from exponential assumptions, alternative modeling approaches (e.g., piecewise or time-series-based) may provide improved accuracy and could represent potential directions for future work. This limitation reflects broader challenges related to cybersecurity data, where empirical indicators such as exploit likelihood are inherently uncertain and may not reliably capture real-world attack dynamics \cite{karaosman2026security}.

\section{Related Work}
\label{sec:related}

The body of evidence~\cite{SLR, 9260512} suggests a growing research interest in the joint analysis of the safety and security of cyber-physical systems, one of which includes self-adaptive systems. This section examines related works as well as their differences to our approach.

\subsection{Safety and Security Modeling Approaches}
Some of the techniques that have been deployed for safety and security modeling include system theory, AFT modeling, and threat and attacker-profile modeling, among others.\\

\noindent\textbf{System Theory Analysis.}
A foundational work in this regard was done by Young and Leveson \cite{10.1145/2556938} wherein they proposed Systems-Theoretic
Accident Model and Processes (STAMP), an accident causality modeling and analysis approach. In their study, security was determined as another property within modern systems. Hence, they argued for the extension of Systems-Theoretic Process Analysis~(STPA), hitherto widely used only for safety analysis, to incorporate security analysis and dubbed it STPA-Sec.
Further to Young and Leveson's study, Pereira et al. \cite{pereira} proposed an integrated approach that combines  STPA and NIST SP800-30, a well-known threats, events and vulnerabilities framework, to automatically analyze and detect conflicts between pairwise reinforcements of various safety and security limitations. However, Pereira et al.'s solution considers safety concerns as a control problem rather than a reliability problem, i.e., the ability to manipulate and direct a system's actions, as opposed to the system's likelihood of experiencing unexpected breakdowns or errors. Furthermore, their solution requires that high-potential risks are known at design time. Although the authors claim that the process of identifying main system risks requires little effort, they discount the need for an advanced understanding of these CPSs and their operational contexts.\\

\noindent\textbf{AT/FT/AFT Generation Approaches.}

Different approaches have been proposed to generate AFT models and are discussed briefly in the ensuing paragraphs. One of the earliest ideas for generating fault- or attack graphs is the one by Swiler et al.~\cite{LauraSwiler2001} wherein they present a tool that generates attack graphs based on an assessment of security attributes and vulnerabilities in computer networks. The authors use vulnerability scanning tools and attack templates to determine network vulnerabilities and introduce configuration files that are similar to the aforementioned dataflow and deployment models. Furthermore, they mention that the integration of an attacker profile might be interesting for determining attack potential. Our implementation differs from Swiler et al.'s because our approach combines the generated \glspl{AT} with \glspl{FT} and additionally, we generate dataflow and deployment models. 

Fovino et al. introduce the concept of integrating the so-called ``attack goal'' resulting from multiple layers of \glspl{AT} in \glspl{FT}. In comparison, Stoelinga et al.~\cite{Stoelinga1} also introduce new model elements to combine \glspl{FT} with \glspl{AT}.
One of the earliest works in \gls{AT} generation, Kotenko et al.~\cite{Kotenko2013} generated an \gls{AT} by using CPEs to identify CVEs for components in which CAPEC metrics are used to generate more complex attack scenarios. However, they focus only on network communication scenarios and thus utilize their network security detection tools to identify possible vulnerabilities. The use of CPEs is similar to our proposed approach, in which \gls{AT} fragments are generated by mining vulnerabilities from CVE databases. However, we utilize software package-related information from a running (ROS) system for the generation of \gls{AT}. In contrast, Ou et al.~\cite{Ou2006} generated attack graphs for network topologies using logic programming. One disadvantage of their approach is that all information from the system must be provided manually in advance by ``facts'' in the logical programming language. \\

\noindent\textbf{Hybrid Analysis Approaches.} Jablonski et al.~\cite{10.1145/3510547.3517922} present a top-down formal method approach based on systems theory for creating attack-defense trees of CPSs by analyzing the system's threat profile for possible risk scenarios depending on its underlying control attributes and communication flows between relevant internal hardware and software sub-components. The systems theoretic approach then aids the objective selection of causal events when utilizing Attack and Fault Tree models. They further demonstrated how the generated attack-defense trees may be reduced to \glspl{AT}, \glspl{FT}, and \glspl{AFT} using a control system case study.\\

\subsection{Runtime Analysis of Self-adaptive Systems}
Studies on runtime monitoring and verification of \glspl{SAS} are not new. In this regard, it is important to mention an earlier study carried out by Villegas et al. \cite{10.1145/1988008.1988020} that highlights the adaptation quality framework, summarised by \cite{tamura:hal-00709943} and provides the relevant background on which further studies such as ours are built upon. With respect to dynamic analysis based on feedback loops in self-adaptive systems, the framework proposed by Klös et al.~\cite{KLOS201828}, leverages adaptation logic by online learning on executable run-time models (RTMs) that capture the system and environment behavior although they do not make use of \glspl{AFT}. In our application of the SAFT-GT pipeline, we generate \glspl{AFT} when a reconfiguration of our ROS-based system starts or ends and then leverage the mean time to failure~(MTTF) value from the Storm \gls{PMC} on \glspl{AFT} (inputted as DFTs (see Section \ref{sec:modelchecking}))  to notify our MAPE-K system of the completion of a reconfiguration of the managed system, which can be considered as a methodological contribution not covered in the work of Klös et al.~\cite{KLOS201828}.

In a previous work, Barbosa et al.~\cite{10.1109/SEAMS.2017.18} presented a tool, Lotus@Runtime which employs a Probabilistic Labelled Transition System (PLTS) to represent the system, and sends a notification when runtime checks of reachability properties of an updated model of a self-adaptive system results in a property violation. Similar to our work, the tool also allows the determination of probabilities and updates them in the model for each transition at runtime. However, one of the limitations of the tool is that the planning and execution phases of the MAPE-K loop are completed outside the LoTuS tool. Similarly, Carwehl et al. \cite{carwehl2023} implemented a runtime verification method that checks the current execution of a body network sensor (BSN) system in the form of a trace against formalized requirements mapped from natural language to Metric Temporal Logic (MTL) properties. In their work, they used adaptation templates contained in a Property Adaptation Pattern (PAP) catalog to automatically instantiate adaptation properties that mirror any requirement changes in the observers of a so-called change manager connected to the managed system. Nonetheless, the monitoring, analysis and planning steps are still coordinated by a \textit{human-in-the-loop} paradigm with the help of a requirements manager. Furthermore, their work did not mention the safety and security concerns of the system under test.

Stemming from the foregoing, our \gls{AFT} generation approach follows the general outline of Steiner et al.~\cite{SteinerLiggesmeyer2016} of combining \glspl{FT} and \glspl{AT} and is in tandem with Fovino et al.'s~\cite{AFTsOrig} work. However, in contrast to their works, our study not only reduces inaccuracies that may occur in manually generated models by making several aspects of the model generation processes semi-/fully automated, such as the generated \glspl{AT}, but our work also facilitates adaptation in \glspl{SAS} by leveraging the outcome of runtime model checking evaluations. An important aspect of our extended work is the automatic conversion of \glspl{AFT} into \glspl{DFT} and the utilization of the \glspl{DFT} for further analysis to determine, on the one hand, the viability of our approach based on the execution time of the EPSS process already explained in detail in Section \ref{sec:modelchecking} and to trigger adaption in a self-adaptive system example. 

\section{Conclusion}
\label{sec:conclusion}

In this work, we presented the SAFESEC toolchain, including all the models and how these models are combined. Furthermore, we extended the toolchain by incorporating it within a \gls{SAS}. Particularly, we used a ROS2-based quadcopter example to demonstrate the application of the extended approach in \glspl{SAS} by integrating the toolchain in the feedback loop of the system and performing time measurements. The process of adaptation in the system is achieved through a probabilistic evaluation of the ``eventuality'' of an attack event with an exponential function obtained from an incorporated Storm PMC within the toolchain.

Then, we carried out an evaluation campaign through expert surveys and discussions during a workshop event. According to the obtained evaluation outcomes, we discussed the strengths and limitations of the toolchain and our extension of the pipeline, as well as possible future research directions. Some of the emerging areas of improvement include the need to make as many parts of the approach automated as possible, considering combining design time and runtime strategies for safety and security to save resources and extending the scope of vulnerabilities (attacks) beyond those that may not be available in public databases. 

In the same vein, we identified that the strong safety backgrounds of the recruited experts in comparison to their security expertise in CPSs may have resulted in key security aspects being missed. As such, the results of our evaluation cannot be generalized. Yet, the results indicate that our approach is both applicable and relevant for the joint analysis of safety and security in complex cyber-physical systems. Finally, we also conducted an experimental evaluation of the extended toolchain that was integrated into the feedback loop of the \gls{SAS}. While we recognize that broader applicability requires further empirical evidence, we consider this an important direction for future work and plan to evaluate the approach across more diverse systems.

\section*{Acknowledgment}
This work was partially supported by the Austrian Science Fund (FWF): I 4701-N and the German Research Foundation (DFG): 435878599. It is also partially supported by Hilti and the Wallenberg AI, Autonomous Systems and Software Program (WASP), funded by the Knut and Alice Wallenberg Foundation.



\bibliographystyle{elsarticle-num}
\bibliography{safesec}

@InProceedings{mauw2006foundations,
author="Mauw, Sjouke
and Oostdijk, Martijn",
editor="Won, Dong Ho
and Kim, Seungjoo",
title="Foundations of Attack Trees",
booktitle="Information Security and Cryptology - ICISC 2005",
year="2006",
publisher="Springer Berlin Heidelberg",
address="Berlin, Heidelberg",
pages="186--198",
isbn="978-3-540-33355-5",
doi="10.1007/11734727_17"
}

@article{schneier1999modeling,
  title={Modeling security threats},
  author={Schneier, Bruce},
  journal={Dr. Dobb's journal},
  volume={24},
  number={12},
  year={1999},
  publisher={Miller Freeman}
}

@ARTICLE{muzammil2024unveiling,
  author={Muzammil, Muteeb Bin and Bilal, Muhammad and Ajmal, Sahar and Shongwe, Sandile C. and Ghadi, Yazeed Y.},
  journal={IEEE Access}, 
  title={Unveiling Vulnerabilities of Web Attacks Considering Man in the Middle Attack and Session Hijacking}, 
  year={2024},
  volume={12},
  number={},
  pages={6365-6375},
  doi={10.1109/ACCESS.2024.3350444}}

@article{lallie2020review,
title = {A review of attack graph and attack tree visual syntax in cyber security},
journal = {Computer Science Review},
volume = {35},
pages = {100219},
year = {2020},
issn = {1574-0137},
doi = {10.1016/j.cosrev.2019.100219},
IGNOREurl = {https://www.sciencedirect.com/science/article/pii/S1574013719300772},
author = {Harjinder Singh Lallie and Kurt Debattista and Jay Bal}
}

@TechReport{vesely1981fault,
  author      = {Vesely, William E and Goldberg, Francine F and Roberts, Norman H and Haasl, David F},
  institution = {Nuclear Regulatory Commission Washington DC},
  title       = {Fault tree handbook},
  year        = {1981},
}

@ARTICLE{DFTs,
  author={Dugan, J.B. and Bavuso, S.J. and Boyd, M.A.},
  journal={IEEE Transactions on Reliability}, 
  title={Dynamic fault-tree models for fault-tolerant computer systems}, 
  year={1992},
  volume={41},
  number={3},
  pages={363-377},
  doi={10.1109/24.159800}

}

@inproceedings{pai2002automatic,
  author={Pai, Ganesh J and Dugan, Joanne Bechta},
  booktitle={13th International Symposium on Software Reliability Engineering, 2002. Proceedings.}, 
  title={Automatic synthesis of dynamic fault trees from UML system models}, 
  year={2002},
  volume={},
  number={},
  pages={243--254},
  doi={10.1109/ISSRE.2002.1173261}
}

@INPROCEEDINGS{1311936,
  author={Raiteri, D.C. and Franceschinis, G. and Iacono, M. and Vittorini, V.},
  booktitle={International Conference on Dependable Systems and Networks, 2004}, 
  title={Repairable fault tree for the automatic evaluation of repair policies}, 
  year={2004},
  volume={},
  number={},
  pages={659-668},
  doi={10.1109/DSN.2004.1311936}}

@article{Stoelinga3,
title = {Fault tree analysis: A survey of the state-of-the-art in modeling, analysis and tools},
journal = {Computer Science Review},
volume = {15-16},
pages = {29-62},
year = {2015},
issn = {1574-0137},
doi = {10.1016/j.cosrev.2015.03.001},
IGNOREurl = {https://www.sciencedirect.com/science/article/pii/S1574013715000027},
author = {Enno Ruijters and Mariëlle Stoelinga},
}

@article{AFTsOrig,
title = {Integrating cyber attacks within fault trees},
journal = {Reliability Engineering \& System Safety},
volume = {94},
number = {9},
pages = {1394-1402},
year = {2009},
note = {ESREL 2007, the 18th European Safety and Reliability Conference},
issn = {0951-8320},
doi = {10.1016/j.ress.2009.02.020},
IGNOREurl = {https://www.sciencedirect.com/science/article/pii/S0951832009000337},
author = {Igor {Nai Fovino} and Marcelo Masera and Alessio {De Cian}}
}

@ARTICLE{1160055,
  author={Kephart, J.O. and Chess, D.M.},
  journal={Computer}, 
  title={The vision of autonomic computing}, 
  year={2003},
  volume={36},
  number={1},
  pages={41-50},
  doi={10.1109/MC.2003.1160055}}

@INPROCEEDINGS{7194653,
  author={Arcaini, Paolo and Riccobene, Elvinia and Scandurra, Patrizia},
  booktitle={2015 IEEE/ACM 10th International Symposium on Software Engineering for Adaptive and Self-Managing Systems}, 
  title={Modeling and Analyzing MAPE-K Feedback Loops for Self-Adaptation}, 
  year={2015},
  volume={},
  number={},
  pages={13-23},
  doi={10.1109/SEAMS.2015.10}}

@InBook{Weyns2013,
  author    = {Weyns, Danny
and Schmerl, Bradley
and Grassi, Vincenzo
and Malek, Sam
and Mirandola, Raffaela
and Prehofer, Christian
and Wuttke, Jochen
and Andersson, Jesper
and Giese, Holger
and G{\"o}schka, Karl M.},
  chapter   = {On Patterns for Decentralized Control in Self-Adaptive Systems},
  editor    = {de Lemos, Rog{\'e}rio
and Giese, Holger
and M{\"u}ller, Hausi A.
and Shaw, Mary},
  pages     = {76--107},
  publisher = {Springer Berlin Heidelberg},
  title     = {Software Engineering for Self-Adaptive Systems II. Lecture Notes in Computer Science},
  year      = {2013},
  address   = {Berlin, Heidelberg},
  isbn      = {978-3-642-35813-5},
  volume    = {7475},
  doi       = {10.1007/978-3-642-35813-5_4},
  urlIGNORE = {https://doi.org/10.1007/978-3-642-35813-5_4},
}

@InProceedings{STA2011,
author="David, Alexandre
and Larsen, Kim G.
and Legay, Axel
and Miku{\v{c}}ionis, Marius
and Poulsen, Danny B{\o}gsted
and van Vliet, Jonas
and Wang, Zheng",
editor="Fahrenberg, Uli
and Tripakis, Stavros",
title="Statistical Model Checking for Networks of Priced Timed Automata",
booktitle="Formal Modeling and Analysis of Timed Systems",
year="2011",
publisher="Springer Berlin Heidelberg",
address="Berlin, Heidelberg",
pages="80--96",
isbn="978-3-642-24310-3",
doi="10.1007/978-3-642-24310-3_7"
}

@article{uppaalSMCtutorial2015,
	title = {Uppaal {SMC} tutorial},
	volume = {17},
	issn = {1433-2787},
	urlIGNORE = {https://doi.org/10.1007/s10009-014-0361-y},
	doi = {10.1007/s10009-014-0361-y},
	number = {4},
	journal = {International Journal on Software Tools for Technology Transfer},
	author = {David, Alexandre and Larsen, Kim G. and Legay, Axel and Mikučionis, Marius and Poulsen, Danny Bøgsted},
	month = aug,
	year = {2015},
	pages = {397--415},
}

@Article{samonas2014cia,
  author  = {Samonas, Spyridon and Coss, David},
  title   = {{{The CIA strikes back: Redefining confidentiality, integrity and availability in security}}},
  number  = {3},
  volume  = {10},
  journal = {Journal of Information System Security},
  year    = {2014},
}

@article{epss2021,
author = {Jacobs, Jay and Romanosky, Sasha and Edwards, Benjamin and Adjerid, Idris and Roytman, Michael},
title = {Exploit Prediction Scoring System (EPSS)},
year = {2021},
issue_date = {September 2021},
publisher = {Association for Computing Machinery},
address = {New York, NY, USA},
volume = {2},
number = {3},
urlIGNORE = {https://doi.org/10.1145/3436242},
doi = {10.1145/3436242},
journal = {Digital Threats},
month = {jul},
articleno = {20},
numpages = {17}
}

@InProceedings{lenin2014attacker,
author="Lenin, Aleksandr
and Willemson, Jan
and Sari, Dyan Permata",
editor="Bernsmed, Karin
and Fischer-H{\"u}bner, Simone",
title="Attacker Profiling in Quantitative Security Assessment Based on Attack Trees",
booktitle="Secure IT Systems",
year="2014",
publisher="Springer International Publishing",
address="Cham",
pages="199--212",
isbn="978-3-319-11599-3",
doi="10.1007/978-3-319-11599-3_12"
}

@INPROCEEDINGS{Stoelinga1,
  author={Kumar, Rajesh and Stoelinga, Mariëlle},
  booktitle={2017 IEEE 18th International Symposium on High Assurance Systems Engineering (HASE)}, 
  title={Quantitative Security and Safety Analysis with Attack-Fault Trees}, 
  year={2017},
  volume={},
  number={},
  pages={25-32},
  doi={10.1109/HASE.2017.12}}

@InProceedings{groner2023model,
author="Groner, Raffaela
and Witte, Thomas
and Raschke, Alexander
and Hirn, Sophie
and Pekaric, Irdin
and Frick, Markus
and Tichy, Matthias
and Felderer, Michael",
editor="Guiochet, J{\'e}r{\'e}mie
and Tonetta, Stefano
and Bitsch, Friedemann",
title="Model-Based Generation of Attack-Fault Trees",
booktitle="Computer Safety, Reliability, and Security (SAFECOMP)",
year="2023",
publisher="Springer Nature Switzerland",
address="Cham",
pages="107--120",
isbn="978-3-031-40923-3",
doi = "10.1007/978-3-031-40923-3_9"
}

@inproceedings{bilge2012before,
author = {Bilge, Leyla and Dumitra\c{s}, Tudor},
title = {Before we knew it: an empirical study of zero-day attacks in the real world},
year = {2012},
isbn = {9781450316514},
publisher = {Association for Computing Machinery},
address = {New York, NY, USA},
urlIGNORE = {https://doi.org/10.1145/2382196.2382284},
doi = {10.1145/2382196.2382284},
booktitle = {Proceedings of the 2012 ACM Conference on Computer and Communications Security},
pages = {833–844},
numpages = {12},
location = {Raleigh, North Carolina, USA},
series = {CCS '12}
}

@Inbook{fournaris2019design,
author="Fournaris, Apostolos P.
and Komninos, Andreas
and Lalos, Aris S.
and Kalogeras, Athanasios P.
and Koulamas, Christos
and Serpanos, Dimitrios",
chapter="Design and Run-Time Aspects of Secure Cyber-Physical Systems",
title="Security and Quality in Cyber-Physical Systems Engineering",
year="2019",
publisher="Springer International Publishing",
address="Cham",
pages="357--382",
isbn="978-3-030-25312-7",
doi="10.1007/978-3-030-25312-7_13",
urlIGNORE="https://doi.org/10.1007/978-3-030-25312-7_13"
}

@inproceedings{braun2023understanding,
author = {Braun, Tobias and Pekaric, Irdin and Apruzzese, Giovanni},
title = {Understanding the Process of Data Labeling in Cybersecurity},
year = {2024},
isbn = {9798400702433},
publisher = {Association for Computing Machinery},
address = {New York, NY, USA},
urlIGNORE = {https://doi.org/10.1145/3605098.3636046},
doi = {10.1145/3605098.3636046},
booktitle = {Proceedings of the 39th ACM/SIGAPP Symposium on Applied Computing},
pages = {1596-1605},
numpages = {10},
location = {Avila, Spain},
series = {SAC '24}
}

@article{ceccarelli2023evaluating,
  title={Evaluating Object (Mis) Detection From a Safety and Reliability Perspective: Discussion and Measures},
  author={Ceccarelli, Andrea and Montecchi, Leonardo},
  journal={IEEE Access},
  year={2023},
  publisher={IEEE},
  volume={11},
  number={},
  pages={44952-44963},
  doi={10.1109/ACCESS.2023.3272979}
}

@article{10.1145/3589227,
author = {Weyns, Danny and Gerostathopoulos, Ilias and Abbas, Nadeem and Andersson, Jesper and Biffl, Stefan and Brada, Premek and Bures, Tomas and Di Salle, Amleto and Galster, Matthias and Lago, Patricia and Lewis, Grace and Litoiu, Marin and Musil, Angelika and Musil, Juergen and Patros, Panos and Pelliccione, Patrizio},
title = {Self-Adaptation in Industry: A Survey},
year = {2023},
issue_date = {June 2023},
publisher = {Association for Computing Machinery},
address = {New York, NY, USA},
volume = {18},
number = {2},
issn = {1556-4665},
urlIGNORE = {https://doi.org/10.1145/3589227},
doi = {10.1145/3589227},
journal = {ACM Trans. Auton. Adapt. Syst.},
month = {may},
articleno = {5},
numpages = {44}
}

@inproceedings{10.1145/3503229.3547048,
author = {Prikler, Liliana Marie and Wotawa, Franz},
title = {Challenges of testing self-adaptive systems},
year = {2022},
isbn = {9781450392068},
publisher = {Association for Computing Machinery},
address = {New York, NY, USA},
urlIGNORE = {https://doi.org/10.1145/3503229.3547048},
doi = {10.1145/3503229.3547048},
booktitle = {Proceedings of the 26th ACM International Systems and Software Product Line Conference - Volume B},
pages = {224–228},
numpages = {5},
location = {Graz, Austria},
series = {SPLC '22}
}

@INPROCEEDINGS{Stoelinga2,
  author={Andr\'{e}, \'{E}tienne and Lime, Didier and Ramparison, Mathias and Stoelinga, Mari\"{e}lle},
  booktitle={2019 19th International Conference on Application of Concurrency to System Design (ACSD)}, 
  title={Parametric Analyses of Attack-Fault Trees}, 
  year={2019},
  volume={},
  number={},
  pages={33-42},
  doi={10.1109/ACSD.2019.00008}
}

@inproceedings{vulner,
  author       = {Irdin Pekaric and
                  Michael Felderer and
                  Philipp Steinm{\"{u}}ller},
  title        = {{VULNERLIZER:} Cross-analysis Between Vulnerabilities and Software
                  Libraries},
  booktitle    = {54th Hawaii International Conference on System Sciences, {HICSS} 2021,
                  Kauai, Hawaii, USA, January 5, 2021},
  pages        = {1--10},
  publisher    = {ScholarSpace},
  year         = {2021},
  urlIGNORE          = {https://hdl.handle.net/10125/71464},
  bibsource    = {dblp computer science bibliography, https://dblp.org}
}

@article{attacks,
title = {A taxonomy of attack mechanisms in the automotive domain},
journal = {Computer Standards \& Interfaces},
volume = {78},
pages = {103539},
year = {2021},
issn = {0920-5489},
doi = {https://doi.org/10.1016/j.csi.2021.103539},
urlIGNORE = {https://www.sciencedirect.com/science/article/pii/S0920548921000349},
author = {Irdin Pekaric and Clemens Sauerwein and Stefan Haselwanter and Michael Felderer}
}

@inproceedings{10.1145/3524844.3528062,
author = {Witte, Thomas and Groner, Raffaela and Raschke, Alexander and Tichy, Matthias and Pekaric, Irdin and Felderer, Michael},
title = {Towards model co-evolution across self-adaptation steps for combined safety and security analysis},
year = {2022},
isbn = {9781450393058},
publisher = {Association for Computing Machinery},
address = {New York, NY, USA},
urlIGNORE = {https://doi.org/10.1145/3524844.3528062},
doi = {10.1145/3524844.3528062},
booktitle = {Proceedings of the 17th Symposium on Software Engineering for Adaptive and Self-Managing Systems},
pages = {106–112},
numpages = {7},
location = {Pittsburgh, Pennsylvania},
series = {SEAMS '22}
}

@InProceedings{ComponentBasedHazard,
author="Giese, Holger
and Tichy, Matthias",
editor="G{\'o}rski, Janusz",
title="Component-Based Hazard Analysis: Optimal Designs, Product Lines, and Online-Reconfiguration",
booktitle="Computer Safety, Reliability, and Security",
year="2006",
publisher="Springer Berlin Heidelberg",
address="Berlin, Heidelberg",
pages="156--169",
doi= "10.1007/11875567_12"
}

@article{macenski2022robot,
author = {Steven Macenski  and Tully Foote  and Brian Gerkey  and Chris Lalancette  and William Woodall },
title = {Robot Operating System 2: Design, architecture, and uses in the wild},
journal = {Science Robotics},
volume = {7},
number = {66},
pages = {eabm6074},
year = {2022},
doi = {10.1126/scirobotics.abm6074},
urlIGNORE = {https://www.science.org/doi/abs/10.1126/scirobotics.abm6074},
eprintIGNORE = {https://www.science.org/doi/pdf/10.1126/scirobotics.abm6074}
}

@article{SLR,
title = {A systematic review on security and safety of self-adaptive systems},
journal = {Journal of Systems and Software},
volume = {203},
pages = {111716},
year = {2023},
issn = {0164-1212},
doi = {https://doi.org/10.1016/j.jss.2023.111716},
urlIGNORE = {https://www.sciencedirect.com/science/article/pii/S0164121223001115},
author = {Irdin Pekaric and Raffaela Groner and Thomas Witte and Jubril Gbolahan Adigun and Alexander Raschke and Michael Felderer and Matthias Tichy}
}

@INPROCEEDINGS{9260512,
  author={Oueidat, Tamara and Flaus, Jean-Marie and Massé, François},
  booktitle={2020 International Conference on Control, Automation and Diagnosis (ICCAD)}, 
  title={A review of combined safety and security risk analysis approaches: Application and Classification}, 
  year={2020},
  volume={},
  number={},
  pages={1-7},
  doi={10.1109/ICCAD49821.2020.9260512}}

@InProceedings{pereira,
author="Pereira, Daniel
and Hirata, Celso
and Pagliares, Rodrigo
and Nadjm-Tehrani, Simin",
editor="Tonetta, Stefano
and Schoitsch, Erwin
and Bitsch, Friedemann",
title="Towards Combined Safety and Security Constraints Analysis",
booktitle="Computer Safety, Reliability, and Security ",
year="2017",
publisher="Springer International Publishing",
address="Cham",
pages="70--80",
isbn="978-3-319-66284-8",
doi="10.1007/978-3-319-66284-8_7"
}

@INPROCEEDINGS{LauraSwiler2001,
  author={Swiler, L.P. and Phillips, C. and Ellis, D. and Chakerian, S.},
  booktitle={Proceedings DARPA Information Survivability Conference and Exposition II. DISCEX'01}, 
  title={Computer-attack graph generation tool}, 
  year={2001},
  volume={2},
  number={},
  pages={307-321 vol.2},
  doi={10.1109/DISCEX.2001.932182}
}

@InProceedings{SteinerLiggesmeyer2016,
  author    = {Max Steiner and Peter Liggesmeyer},
  booktitle = {Workshop DECS (ERCIM/EWICS workshop on dependable embedded and cyber-physical systems) of the 32nd international conference on computer safety (SAFECOMP)},
  title     = {Combination of Safety and Security Analysis - Finding Security Problems That Threaten the Safety of a System},
  year      = {2013},
}

@INPROCEEDINGS{Kotenko2013,
  author={Kotenko, Igor and Chechulin, Andrey},
  booktitle={2013 5th International Conference on Cyber Conflict (CYCON 2013)}, 
  title={A Cyber Attack Modeling and Impact Assessment framework}, 
  year={2013},
  volume={},
  number={},
  pages={1-24},
  doi={}
}

@inproceedings{pekaric2023streamlining,
    author       = {Irdin Pekaric and
                  Markus Frick and
                  Jubril Gbolahan Adigun and
                  Raffaela Groner and
                  Thomas Witte and
                  Alexander Raschke and
                  Michael Felderer and
                  Matthias Tichy},
    editor       = {Tung X. Bui},
    title        = {Streamlining Attack Tree Generation: {A} Fragment-Based Approach},
    booktitle    = {57th Hawaii International Conference on System Sciences, {HICSS} 2024,
                  Hilton Hawaiian Village Waikiki Beach Resort, Hawaii, USA, January
                  3-6, 2024},
    pages        = {7447--7456},
    publisher    = {ScholarSpace},
    year         = {2024},
    urlIGNORE          = {https://hdl.handle.net/10125/107280},
    bibsource    = {dblp computer science bibliography, https://dblp.org}
}

@inproceedings{Ou2006,
    author = {Ou, Xinming and Boyer, Wayne F. and McQueen, Miles A.},
    title = {A scalable approach to attack graph generation},
    year = {2006},
    isbn = {1595935185},
    publisher = {Association for Computing Machinery},
    address = {New York, NY, USA},
    urlIGNORE = {https://doi.org/10.1145/1180405.1180446},
    doi = {10.1145/1180405.1180446},
    booktitle = {Proceedings of the 13th ACM Conference on Computer and Communications Security},
    pages = {336–345},
    numpages = {10},
    location = {Alexandria, Virginia, USA},
    series = {CCS '06}
}

@inproceedings{10.1145/3510547.3517922,
    author = {Jablonski, Matthew and Wijesekera, Duminda and Singhal, Anoop},
    title = {Generating Cyber-Physical System Risk Overlays for Attack and Fault Trees using Systems Theory},
    year = {2022},
    isbn = {9781450392297},
    publisher = {Association for Computing Machinery},
    address = {New York, NY, USA},
    urlIGNORE = {https://doi.org/10.1145/3510547.3517922},
    doi = {10.1145/3510547.3517922},
    booktitle = {Proceedings of the 2022 ACM Workshop on Secure and Trustworthy Cyber-Physical Systems},
    pages = {13–20},
    numpages = {8},
    location = {Baltimore, MD, USA},
    series = {Sat-CPS '22}
}

@article{KLOS201828,
    title = {Comprehensible and dependable self-learning self-adaptive systems},
    journal = {Journal of Systems Architecture},
    volume = {85-86},
    pages = {28-42},
    year = {2018},
    issn = {1383-7621},
    doi = {https://doi.org/10.1016/j.sysarc.2018.03.004},
    urlIGNORE = {https://www.sciencedirect.com/science/article/pii/S1383762117304472},
    author = {Verena Klös and Thomas Göthel and Sabine Glesner}
}

@InProceedings{10.1007/978-3-642-33678-2_9,
    author="Edifor, Ernest
    and Walker, Martin
    and Gordon, Neil",
    editor="Ortmeier, Frank
    and Daniel, Peter",
    title="Quantification of Priority-OR Gates in Temporal Fault Trees",
    booktitle="Computer Safety, Reliability, and Security",
    year="2012",
    publisher="Springer Berlin Heidelberg",
    address="Berlin, Heidelberg",
    pages="99--110",
    isbn="978-3-642-33678-2",
doi={10.1007/978-3-642-33678-2_9}
}

@Article{gherardi2013variability,
    author    = {Gherardi, Luca},
    title     = {{{Variability modeling and resolution in component-based robotics systems}}},
    journal   = {Ph. D. Thesis},
    publisher = {Universit{\`a} degli Studi di Bergamo Bergamo, Italy},
    year      = {2013}
}

@inproceedings{10.1145/1988008.1988020,
    author = {Villegas, Norha M. and M\"{u}ller, Hausi A. and Tamura, Gabriel and Duchien, Laurence and Casallas, Rubby},
    title = {A framework for evaluating quality-driven self-adaptive software systems},
    year = {2011},
    isbn = {9781450305754},
    publisher = {Association for Computing Machinery},
    address = {New York, NY, USA},
    urlIGNORE = {https://doi.org/10.1145/1988008.1988020},
    doi = {10.1145/1988008.1988020},
    booktitle = {Proceedings of the 6th International Symposium on Software Engineering for Adaptive and Self-Managing Systems},
    pages = {80–89},
    numpages = {10},
    location = {Waikiki, Honolulu, HI, USA},
    series = {SEAMS '11}
}

@incollection{tamura:hal-00709943,
  TITLE = {{Towards Practical Runtime Verification and Validation of Self-Adaptive Software Systems}},
  AUTHOR = {Tamura, Gabriel and Villegas, Norha and M{\"u}ller, Hausi and P. Sousa, Jo{\~a}o and Becker, Basil and Pezz{\`e}, Mauro and Karsai, Gabor and Mankovskii, Serge and Sch{\"a}fer, Wilhelm and Tahvildari, Ladan and Wong, Kenny},
  urlIGNORE = {https://inria.hal.science/hal-00709943},
  BOOKTITLE = {{Software Engineering for Self-Adaptive Systems 2}},
  EDITOR = {Rogerio de Lemos and Holger Giese and Hausi M{\"u}ller and Mary Shaw},
  PUBLISHER = {{Springer}},
  SERIES = {LNCS},
  VOLUME = {7475},
  PAGES = {116-141},
  YEAR = {2012},
  doi = {10.1007/978-3-642-35813-5_5},
  MONTH = Jun,
  HAL_ID = {hal-00709943},
  HAL_VERSION = {v1},
}

@article{10.1145/2556938,
author = {Young, William and Leveson, Nancy G.},
title = {An integrated approach to safety and security based on systems theory},
year = {2014},
issue_date = {February 2014},
publisher = {Association for Computing Machinery},
address = {New York, NY, USA},
volume = {57},
number = {2},
issn = {0001-0782},
urlIGNORE = {https://doi.org/10.1145/2556938},
doi = {10.1145/2556938},
journal = {Commun. ACM},
month = feb,
pages = {31–35},
numpages = {5}
}

@article{MENS2006125,
	title = {A Taxonomy of Model Transformation},
	volume = {152},
	issn = {1571-0661},
	urlIGNORE = {https://www.sciencedirect.com/science/article/pii/S1571066106001435},
	doi = {10.1016/j.entcs.2005.10.021},
	series = {Proceedings of the International Workshop on Graph and Model Transformation ({GraMoT} 2005)},
	pages = {125--142},
	journal = {Electronic Notes in Theoretical Computer Science},
	journaltitle = {Electronic Notes in Theoretical Computer Science},
	shortjournal = {Electronic Notes in Theoretical Computer Science},
	author = {Mens, Tom and Van Gorp, Pieter},
	urldate = {2023-11-27},
	date = {2006-03-27},
	year={2006},
}

@inproceedings{Galileo2000,
author = {Coppit, David and Sullivan, Kevin J.},
title = {Galileo: a tool built from mass-market applications},
year = {2000},
isbn = {1581132069},
publisher = {Association for Computing Machinery},
address = {New York, NY, USA},
urlIGNORE = {https://doi.org/10.1145/337180.337622},
doi = {10.1145/337180.337622},
booktitle = {Proceedings of the 22nd International Conference on Software Engineering},
pages = {750–753},
numpages = {4},
location = {Limerick, Ireland},
series = {ICSE '00}
}

@InProceedings{Stoelinga4,
author="Budde, Carlos E.
and Kolb, Christina
and Stoelinga, Mari{\"e}lle",
editor="Abate, Alessandro
and Marin, Andrea",
title="Attack Trees vs. Fault Trees: Two Sides of the Same Coin from Different Currencies",
booktitle="Quantitative Evaluation of Systems",
year="2021",
publisher="Springer International Publishing",
address="Cham",
pages="457--467",
isbn="978-3-030-85172-9",
doi="10.1007/978-3-030-85172-9_24"
}

@article{arazzi2023nlp,
  author       = {Marco Arazzi and
                  Dincy R. Arikkat and
                  Serena Nicolazzo and
                  Antonino Nocera and
                  Rafidha Rehiman K. A. and
                  P. Vinod and
                  Mauro Conti},
  title        = {{NLP}-Based Techniques for Cyber Threat Intelligence},
  journal      = {CoRR},
  volume       = {abs/2311.08807},
  year         = {2023},
  urlIGNORE          = {https://doi.org/10.48550/arXiv.2311.08807},
  doi          = {10.48550/ARXIV.2311.08807},
  eprinttypeIGNORE    = {arXiv},
  eprintIGNORE       = {2311.08807},
  bibsource    = {dblp computer science bibliography, https://dblp.org}
}

@INPROCEEDINGS{DFTsUncovered,
  author={Junges, Sebastian and Guck, Dennis and Katoen, Joost-Pieter and Stoelinga, Mariëlle},
  booktitle={2016 46th Annual IEEE/IFIP International Conference on Dependable Systems and Networks (DSN)}, 
  title={Uncovering Dynamic Fault Trees}, 
  year={2016},
  volume={},
  number={},
  pages={299-310},
  doi={10.1109/DSN.2016.35}}

@INPROCEEDINGS {carwehl2023,
author = { Carwehl, Marc and Vogel, Thomas and Rodrigues, Genaina Nunes and Grunske, Lars },
booktitle = { 2023 IEEE/ACM 18th Symposium on Software Engineering for Adaptive and Self-Managing Systems (SEAMS) },
title = {{ Runtime Verification of Self-Adaptive Systems with Changing Requirements }},
year = {2023},
volume = {},
ISSN = {},
pages = {104-114},
doi = {10.1109/SEAMS59076.2023.00024},
urlIGNORE = {https://doi.ieeecomputersociety.org/10.1109/SEAMS59076.2023.00024},
publisher = {IEEE Computer Society},
address = {Los Alamitos, CA, USA},
month = {May}}

@inproceedings{10.1109/SEAMS.2017.18,
author = {Barbosa, Davi Monteiro and de Moura Lima, R\^{o}mulo Gadelha and Maia, Paulo Henrique Mendes and Junior, Evil\'{a}sio Costa},
title = {Lotus@Runtime: a tool for runtime monitoring and verification of self-adaptive systems},
year = {2017},
isbn = {9781538615508},
publisher = {IEEE Press},
urlIGNORE = {https://doi.org/10.1109/SEAMS.2017.18},
doi = {10.1109/SEAMS.2017.18},
booktitle = {Proceedings of the 12th International Symposium on Software Engineering for Adaptive and Self-Managing Systems},
pages = {24–30},
numpages = {7},
location = {Buenos Aires, Argentina},
series = {SEAMS '17}
}

@INPROCEEDINGS{9224539,
  author={Ponsard, Christophe and Deprez, Jean-Christophe and Darimont, Robert},
  booktitle={2020 IEEE Workshop on Formal Requirements (FORMREQ)}, 
  title={Formalizing Security and Safety Requirements by Mapping Attack-Fault Trees on Obstacle Models with Constraint Programming Semantics}, 
  year={2020},
  volume={},
  number={},
  pages={8-13},
  doi={10.1109/FORMREQ51202.2020.00009}}

@article{kriaa_pietre-cambacedes_bouissou_halgand_2015, title={A survey of approaches combining safety and security for industrial control systems}, volume={138}, ISSN={0951-8320}, DOI={10.1016/j.ress.2015.02.008}, abstractNote={The migration towards digital control systems creates new security threats that can endanger the safety of industrial infrastructures. Addressing the convergence of safety and security concerns in this context, we provide a comprehensive survey of existing approaches to industrial facility design and risk assessment that consider both safety and security. We also provide a comparative analysis of the different approaches identified in the literature. - Highlights: • We raise awareness of safety and security convergence in numerical control systems. • We highlight safety and security interdependencies for modern industrial systems. • We give a survey of approaches combining safety and security engineering. • We discuss the potential of the approaches to model safety and security interactions}, journal={Reliability Engineering and System Safety}, author={Kriaa, Siwar and Pietre-Cambacedes, Ludovic and Bouissou, Marc and Halgand, Yoran}, year={2015}, month={Jul}, pages={p. 156–178} }

@inproceedings{bloomfield2013security,
  title={Security-informed safety: if it’s not secure, it’s not safe},
  author={Bloomfield, Robin and Netkachova, Kateryna and Stroud, Robert},
  booktitle={International workshop on software engineering for resilient systems},
  pages={17--32},
  year={2013},
  organization={Springer}
}

@article{fovino2009integrating,
  title={Integrating cyber attacks within fault trees},
  author={Fovino, Igor Nai and Masera, Marcelo and De Cian, Alessio},
  journal={Reliability Engineering \& System Safety},
  volume={94},
  number={9},
  pages={1394--1402},
  year={2009},
  publisher={Elsevier}
}

@article{sauerwein2019analysis,
  title={An analysis and classification of public information security data sources used in research and practice},
  author={Sauerwein, Clemens and Pekaric, Irdin and Felderer, Michael and Breu, Ruth},
  journal={Computers \& security},
  volume={82},
  pages={140--155},
  year={2019},
  publisher={Elsevier}
}

@inproceedings{karaosman2026security,
author = {Karaosman, Emir and Rizvani, Advije and Pekaric, Irdin},
title = {{Security Barriers to Trustworthy AI-Driven Cyber Threat Intelligence in Finance: Evidence from Practitioners}},
booktitle = {Proceedings of the Sixteenth ACM Conference on Data and Application Security and Privacy (CODASPY)},
year = {2026},
location = {Frankfurt am Main, Germany},
doi = {10.1145/3800506.3803505}
}

@inproceedings{pfister2025department,
  title={Department-Specific Security Awareness Campaigns: A Cross-Organizational Study of HR and Accounting},
  author={Pfister, Matthias and Apruzzese, Giovanni and Pekaric, Irdin},
  booktitle={2025 APWG Symposium on Electronic Crime Research (eCrime)},
  pages={1--17},
  year={2025},
  organization={IEEE}
}





\end{document}